\documentclass{emulateapj}
\usepackage{apjfonts}
\usepackage{epsfig,color}
\usepackage{mathrsfs,amsmath}
\usepackage{ulem}
\usepackage{setspace}

\def\sss{\scriptscriptstyle}
\def\bhm{M_{\bullet}}
\def\bhl{L_{\bullet}}

\def\dotm{\dot{m}}
\def\mdot{\dot{m}}
\def\Mdot{\dot{M}_{\bullet}}

\def\ellkappa{\ell_{\kappa}}
\def\ergs{\rm erg~s^{-1}}
\def\etal{{et al.}}
\def\fblr{f_{\sss\rm BLR}}

\def\feii{Fe {\sc ii}}

\def\Gammax{\Gamma_{_X}}
\def\kms{\rm km~s^{-1}}
\def\kappabol{\kappa_{_{\rm Bol}}}

\def\mathdotM{\dot{\mathscr{M}}}
\def\mdotmin{\dot{m}_{\rm min}}

\def\oiii{[O\,{\sc iii}]}

\def\pp{\prime\prime}

\def\rblr{R_{\sss{\rm BLR}}}
\def\sunm{M_\odot}

\def\dbh{\mathscr{D}_{\bullet}}
\def\dl{\mathscr{D}_{\rm L}}

\def\taublr{\tau_{_{\rm BLR}}}

\def\mgii{Mg~{\sc ii}}

\def\Dbh{\mathscr{D}_{\bullet}}
\def\Dl{\mathscr{D}_{\rm L}}
\def\dbh{d_{\bullet}}
\def\dl{d_{\rm L}}
\def\lk{\ell_{\kappa}}
\def\ud{\mathrm{d}}
\def\dli{d^{i}_{\rm L}}
\def\dbhu{d_{\epsilon}}

\journalinfo{The Astrophysics Journal, in press}
\slugcomment{Received 2014 January 16;  accepted 2014 July 29}

\begin{document}

\title{Supermassive black holes with high accretion rates in active galactic nuclei: \\
II. the most luminous standard candles in the Universe}

\author
{Jian-Min Wang\altaffilmark{1,5},
Pu Du\altaffilmark{1},
Chen Hu\altaffilmark{1},
Hagai Netzer\altaffilmark{2},
Jin-Ming Bai\altaffilmark{3},
Kai-Xing Lu\altaffilmark{4},\\
Shai Kaspi\altaffilmark{2},
Jie Qiu\altaffilmark{1},
Yan-Rong Li\altaffilmark{1} and
Fang Wang\altaffilmark{3}\\
(SEAMBH collaboration)}

\altaffiltext{1}
{Key Laboratory for Particle Astrophysics, Institute of High Energy Physics,
Chinese Academy of Sciences, 19B Yuquan Road, Beijing 100049, China}

\altaffiltext{2}
{Wise Observatory, School of Physics and Astronomy, Tel-Aviv University, Tel-Aviv 69978, Israel}

\altaffiltext{3}
{Yunnan Observatory, Chinese Academy of Sciences, Kunming 650011, Yunnan, China}

\altaffiltext{4}
{Astronomy Department, Beijing Normal University, Beijing 100875, China}

\altaffiltext{5}
{National Astronomical Observatories of China, Chinese Academy of Sciences,
 20A Datun Road, Beijing 100020, China}

\begin{abstract}
This is the second in a series of papers reporting on a large reverberation
mapping (RM) campaign to measure black hole (BH) mass in high accretion rate
active galactic nuclei (AGNs). The goal is to identify super-Eddington accreting
massive black holes (SEAMBHs) and to use their unique properties to construct
a new method for measuring cosmological distances. 
Based on theoretical models, the saturated bolometric luminosity of such sources
is proportional to the BH mass which can be used to obtain their distance. 
Here we report on five new RM measurements and show that
in four of the cases we can measure the BH mass and three of these sources are 
SEAMBHs. Together with the three sources from our earlier work, we now have six 
new sources of this type. We use a novel method based on a minimal radiation 
efficiency  to identify  nine additional SEAMBHs from earlier RM-based mass 
measurements. We use a Bayesian analysis to determine the parameters of the 
new distance expression, and the method uncertainties, from the observed properties 
of the objects in the sample. The ratio of the newly measured distances
to the standard cosmological ones has a mean scatter of 0.14 dex, indicating 
that SEAMBHs can be use as cosmological distance probes.
With their high luminosity, long period of activity and large
numbers at high redshifts, SEAMBHs have a potential to extend  the cosmic
distance ladder beyond the range now explored by type Ia supernovae.
\end{abstract}


\section{Introduction}
The powerful emission of active galactic nuclei (AGN) is thought to originate 
from gas accretion onto supermassive black holes (BHs). In particular, a large 
fraction of type-I (unobscured) AGNs are  powered by super-Eddington accreting 
massive black holes (SEAMBHs) that are characterised by a large Eddington ratio,
$L_{\rm Bol}/L_{\rm Edd} \gtrsim 1$, where $L_{\rm Bol}$ is the bolometric 
luminosity and $L_{\rm Edd}$ is the Eddington luminosity (Nobuta et al. 2012; 
Kelly \& Shen 2013; Netzer \& Trakhtenbrot 2014). In the local Universe, most 
objects of this type are classified as narrow line Seyfert 1 galaxies 
(NLS1s). They have prominent features of relatively narrow broad emission lines, 
strong \feii\, lines, weak \oiii\ lines, and very strong soft X-ray continuum 
(Osterbrock \& Pogge 1985; Boller \etal\, 1996). SEAMBHs as central engines of 
NLS1s are not entirely understood. It is thought 
that they contain slim accretion disks where photon trapping is very important 
(e.g., Abramowicz \etal\, 1988; Wang \& Zhou 1999a; Ohsuga \etal\, 2001).

The accelerated  expansion of the Universe has been extensively studied using type 
Ia supernovae (SNe-Ia) (Riess \etal\, 1998; Perlmutter \etal\, 1999). Much attention 
has been given to improve the measurements of SN-Ia in two directions: 1) more detections 
of SN-Ia at high redshifts and 2) high accuracy of SN-Ia calibrated using Cepheids 
(Freedman \& Madore 2010; Riess \etal\, 2011; Freedman \& Madore 2013). This method 
based on SN-Ia, however, is limited to redshifts up to about $z\sim 1.5$, even for the next 
generation of extremely large telescopes both on the ground or in space (Hook 2012). 
Thus the exact dynamical state of the Universe at higher redshifts is poorly known 
because of the lack of reliable luminous candles (Weinberg \etal\, 2013, Kim et al. 
2013). 

The very high luminosity of AGNs makes them observable to very large distances, up 
to a redshift of about 7 (Mortlock et al. 2011). 
In principle, they could be used to follow the expansion 
of the early Universe, beyond the distance accessible by the SNe-Ia method. 
However, the diversity in their intrinsic properties (Ho 2008; Netzer 2013) hampers 
the use of most AGNs as  distance indicators. There were several recent attempts, 
based on reverberation mapping (RM) of dust and gas near the central BH, to test  
the idea of using AGNs as standard cosmological candles.
One of them makes use of the known correlation between the broad line region size 
($\rblr$) and the continuum luminosities at 5100\AA\, ($L_{5100}$) found in $\sim 50$ 
AGNs (Kaspi \etal\, 2000; Bentz \etal\, 2013; Peterson 2013). Since $\rblr$ is empirically 
determined by measuring the time lag between the  H$\beta$ line and the visual continuum, 
it was suggested to derive this cosmologically independent size by monitoring a large number 
of AGNs at various redshifts (Horne \etal\, 2003; Teerikorpi 2011; Watson \etal\, 2011; 
Czerny \etal\, 2013; Bentz \etal\, 2013; Melia 2014,
Elvis \& Karovska 2002). X-ray variability correlated
with BH mass or luminosity has also been suggested to estimate cosmic distances 
(La Franca et al. 2014). Similarly, dust RM (correlating the variable rest frame $V$ 
and $K$ magnitudes) was suggested to measure the innermost size of the dusty ``torus'' 
as an alternative size measure (Hoenig 2014; Yoshii et al. 2014). Other 
suggestions involve radio megamasers  (Humphreys et al. 2013), and other AGN components
(see Marziani \& Sulentic 2013 for a review of such methods).

The goal of this paper is to show that SEAMBHs can be used as new probes of cosmological 
distances provided their mass is directly measured by methods such as RM. These objects 
show a unique dependence of their bolometric luminosity on the BH mass ($\bhm$) which, 
given a proper calibration, can be used to infer cosmological distances. This idea was 
first introduced by Wang et al. (2013, hereafter W13) who showed that selecting SEAMBHs 
by the slope of their X-ray spectral energy distribution (SED, Wang et al. 2004), and 
estimating their BH 
mass using the $\rblr-L_{5100}$ relation (e.g., Kaspi et al. 2000, 2005, Bentz et al. 
2013) can be used to isolate a sub-group of such sources whose properties are suitable 
for measuring cosmological distances. However,  W13 could not test their idea directly 
since accurate masses of SEAMBHs were not available at the time.

In 2012 we started a large RM campaign to measure BH mass 
in SEAMBH candidates and to identify a large enough number of such sources that can 
be used to establish the method and to calibrate them as standard cosmological candles.
The first paper in the series (Du et al. 2014, hereafter Paper I) described  
our observing project that was carried out in Lijiang, China. Paper I provides detailed 
information about the sample selection, telescope and spectrograph, the light curves 
and cross correlation (CC) analysis, and the mass measurements. The observations reported 
in the present paper were obtained during the same observing season reported in paper-I. 
Their analysis revealed the presence of of three confirmed SEAMBHs and one source which probably
belongs to this group but with only an upper limit on the BH mass..
Combining the new confirmed SEAMBHs 
with the three objects reported in paper-I, and with several other SEAMBHs identified in 
earlier RM experiments, results in a sample which is large enough to test the idea
that such objects can be used as standard cosmological candles. 

The paper is arranged as follows: \S2 describes various types of accretion disks in AGNs 
and gives the necessary approximations to estimate the accretion rate. \S3 presents new 
RM observations and their analysis. In \S4 we list the newly obtained mass and accretion 
rates, and explain our unique method of identifying SEAMBHs. \S5 gives a full description 
of the new method to measure distances with SEAMBHs by way of a rigorous analysis of the 
errors associated with the method. In the last section we draw some conclusions regarding 
the merit of the new method and how can we improve it in the future.

\section{AGN accretion disks}
We consider three main types of accretion flows with angular momentum onto BHs.  
The properties of all such flows are determined by the dimensionless accretion rate,
$\dotm=\eta\mathdotM$,
where
$\mathdotM=\Mdot c^2/L_{\rm Edd}$, 
$\eta$ is the mass-to-radiation conversion 
efficiency and $\Mdot$ the mass accretion rate. Here $\dotm$ is equivalent to the 
Eddington ratio defined by $L_{\rm Bol}/L_{\rm Edd}$, where 
$L_{\rm Edd}=1.5\times 10^{38}\left(\bhm/\sunm\right)\ergs$ for solar composition gas.

For  $\mathdotM\ll 1$, the accretion flow becomes advection-dominated in the radial 
direction and radiative cooling is inefficient and dominated by optically thin free-free 
emission (Narayan \& Yi 1994). This situation probably applies to LINERs (Ho 2008). 
For small to moderate $\mathdotM$, the flow can be described as optically thick 
geometrically thin accretion disk, with $H/R\ll 1$, where $H$ is the height of the 
disk at a radius $R$ (Shakura \& Sunyaev 1973; hereafter SS73). In such cases, the 
radiative efficiency $\eta$ depends on the radius of the last stable orbit which is 
determined by the BH spin.

Standard optically thick geometrically thin accretion disk models assume Keplerian 
rotation in almost perfect circular orbits and very slow inward drift velocity. 
Blackbody emission is very efficient at all radii and the dissipated energy is 
released locally (see SS73). 
While recent studies of thin disks show that the SED can differ, substantially,
from the SS73 approximation (due to e.g., Comptonization, radiative transfer in the 
atmosphere, etc; Frank \etal\, 2002, Kato \etal\, 2003), none of these effects changes 
significantly the total energy released locally by the disk.

The SS73 model has a canonical spectrum whose (BH mass dependent) low frequency SED is given 
approximately by $F_{\nu}\propto \nu^{1/3}$. Over this part of the spectrum one can use the 
standard disk equations to estimate the mass accretion rate, 
$\Mdot=0.53~\left(l_{44}/\cos i\right)^{3/2}m_7^{-1}~M_{\odot}{\rm yr^{-1}}$ 
(see e.g., Frank \etal\, 2002; Netzer 2013), 
where $i$ is the inclination angle of the accretion disk to the line of sight,
$l_{44}=4\pi \dl^2 (\lambda F_{\lambda})/10^{44}\ergs$ and $F_{\lambda}$ is the observed flux 
at $\lambda=5100(1+z)$\AA. The corresponding 
$\dotm$ by the following expressions (slightly adopted from earlier works by using our preferred 
wavelength of 5100\AA),
\begin{equation}
\dotm_{_{\rm SS}}=20.1\left(\frac{l_{44}}{\cos i}\right)^{3/2}m_7^{-2}\eta \, .
\label{eq:SS_2}
\end{equation}
We use $\dotm_{_{\rm SS}}$ to refer to $\dot{m}$ derived in this way (Collin et al. 2002)
and assume an averaged $\cos i\approx 0.75$ for type-I AGN. 
Obviously $\dotm_{_{\rm SS}}$ and $\Mdot$ require the knowledge of the distance 
to the object (the quantity we want to determine in this paper). This however is 
only required to justify that the objects we are selecting, at small redshift, are 
indeed SEAMBHs and the associated uncertainties do not affect much the final results since 
the derived distances are insensitive to the exact values of $\dotm$ (see details below).

\begin{deluxetable*}{lllllcrrc}
\tablecolumns{7}
\tablewidth{0pt}
\setlength{\tabcolsep}{4pt}
\tablecaption{The Lijiang project: targets and observations}
\tabletypesize{\scriptsize}
\tablehead{
\colhead{Object}                &
\colhead{$\alpha_{2000}$}       &
\colhead{$\delta_{2000}$}       &
\colhead{redshift}              &
\colhead{monitoring period}     &
\colhead{$N_{\rm spec}$}                   &
\multicolumn{2}{c}{Comparison stars} &
\colhead{Note on $\tau_{_{\rm BLR}}$ }                  \\ \cline{7-8}
\colhead{}                      &
\colhead{}                      &
\colhead{}                      &
\colhead{}                      &
\colhead{}                      &
\colhead{}                      &
\colhead{$R_*$}                 &
\colhead{P.A.}                  &
\colhead{}
}
\startdata
Mrk 335          & 00 06 19.5 & $+$20 12 10 & 0.0258 & Oct., 2012 $-$ Feb., 2013& 91& $80^{\pp}.7$  & $174.5^{\circ}$ &Yes \\
Mrk 1044         & 02 30 05.5 & $-$08 59 53 & 0.0165 & Oct., 2012 $-$ Feb., 2013& 77& $207^{\pp}.0$ &$-143.0^{\circ}$&Yes \\
IRAS 04416+1215  & 04 44 28.8 & $+$12 21 12 & 0.0889 & Oct., 2012 $-$ Mar., 2013& 92& $137^{\pp}.9$ &$-55.0^{\circ}$ &No \\
Mrk 382          & 07 55 25.3 & $+$39 11 10 & 0.0337 & Oct., 2012 $-$ May., 2013&123& $198^{\pp}.4$ & $-24.6^{\circ}$ &Yes\\
Mrk 142          & 10 25 31.3 & $+$51 40 35 & 0.0449 & Nov., 2012 $-$ Apr., 2013&119& $113^{\pp}.1$ & $155.2^{\circ}$ &Yes\\
MCG $+06-26-012$ & 11 39 13.9 & $+$33 55 51 & 0.0328 & Jan., 2013 $-$ Jun., 2013& 34& $204^{\pp}.3$ & $46.1^{\circ}$ &Yes\\
IRAS F12397+3333 & 12 42 10.6 & $+$33 17 03 & 0.0435 & Jan., 2013 $-$ May., 2013& 51& $189^{\pp}.0$ & $130.0^{\circ}$ &Yes\\
Mrk 42           & 11 53 41.8 & $+$46 12 43 & 0.0246 & Jan., 2013 $-$ Apr., 2013& 53& $234^{\pp}.4$ & $33.8^{\circ}$ &No\\
Mrk 486          & 15 36 38.3 & $+$54 33 33 & 0.0389 & Mar., 2013 $-$ Jul., 2013& 45& $193^{\pp}.8$ &$-167.0^{\circ}$&Yes \\
Mrk 493          & 15 59 09.6 & $+$35 01 47 & 0.0313 & Apr., 2013 $-$ Jun., 2013& 27& $155^{\pp}.3$ & $98.5^{\circ}$ &Yes/No
\enddata
\tablecomments{We include the three objects reported in paper I.
$N_{\rm spec}$ is the numbers of spectroscopic epochs, $R_*$ is the angular distance
between the object and the comparison star and PA is the position angle from the AGN to the
comparison star. The last column contains notes on the H$\beta$ time lags: ``Yes'' means 
significant lag and ``No" lag could
not be measured. The special case of Mrk\,493 is explained in the text of the paper.
}
\end{deluxetable*}

The third type of flow is the one with large $\mathdotM$. In this case, radiation pressure 
dominates the flow geometry at almost all radii and the disk becomes  slim or thick with 
$H \lesssim R$. The concept of slim disks was originally suggested by Paczynsky \& Bisnovatyi-Kogan 
(1981) and Muchotrzeb \& Paczynski (1982) to get rid of the singularity of gas density at the inner 
edge of the SS73 disks. Such systems have  been extensively studied using vertically-averaged 
equations (Matsumoto \etal\, 1984; Muchotrzeb-Czerny 1986). Abramowicz \etal\, (1988) used
the equations to treat slim disks as transonic flows of super-Eddington accretion onto black 
holes. Some of the recent studies include include  Szuszkiewicz \etal\, (1996), Beloborodov 
(1998), Wang \& Zhou (1999a,b), Fukue (2000), Mineshige \etal\, (2000), Watarai \& Mineshige 
(2001, 2013), Sadowski \etal\, (2011, 2013) and McKenney et al. (2013). Slim disks are 
characterized by sub-Keplerian 
rotation and transonic radial motion, which results in a non-localized energy conservation.
These three properties are very  different from the SS73 disks and there 
is no simple analytical solution except for the case of extremely high accretion 
rates, which can be described by a self-similar solution (Wang \& Zhou 1999a; Wang \& Netzer 2003).

As a result of the fast radial transportation in slim disks,  most photons will be trapped and 
advected into the  BH before escaping. This results in  
inefficient emission of radiation and a big reduction in $\eta$.
Some models suggest that in this case, $\eta\propto \mathdotM^{-1}$ (Wang \& Zhou 1999a; Mineshige 
\etal\, 2000; Sadowski et al. 2011). This leads to the so called ``saturated luminosity'' 
given by:
\begin{equation}
\bhl=\ell_0\bhm \, ,
\label{eq:bhl}
\end{equation}
where $\ell_0= 3.0\times 10^{38}[1+\ln(\mathdotM/50)]~{\ergs}M_{\odot}^{-1}$ 
(Wang \& Zhou 1999a; Mineshige et al. 2000). Thus $\bhl\approx 2L_{\rm Edd}$ over a large range 
of accretion rates around $\mathdotM=50$. Given an accurately measured BH mass, one can deduce 
$\bhl$ and hence the BH distance provided there is a way (i.e., a reliable bolometric correction 
factor) to convert $\bhl$ to the monochromatic luminosity of the disk at an accessible wavelength 
(see \S5).

\begin{figure*}[t!]
\begin{center}
\includegraphics[angle=0,width=0.9\textwidth]{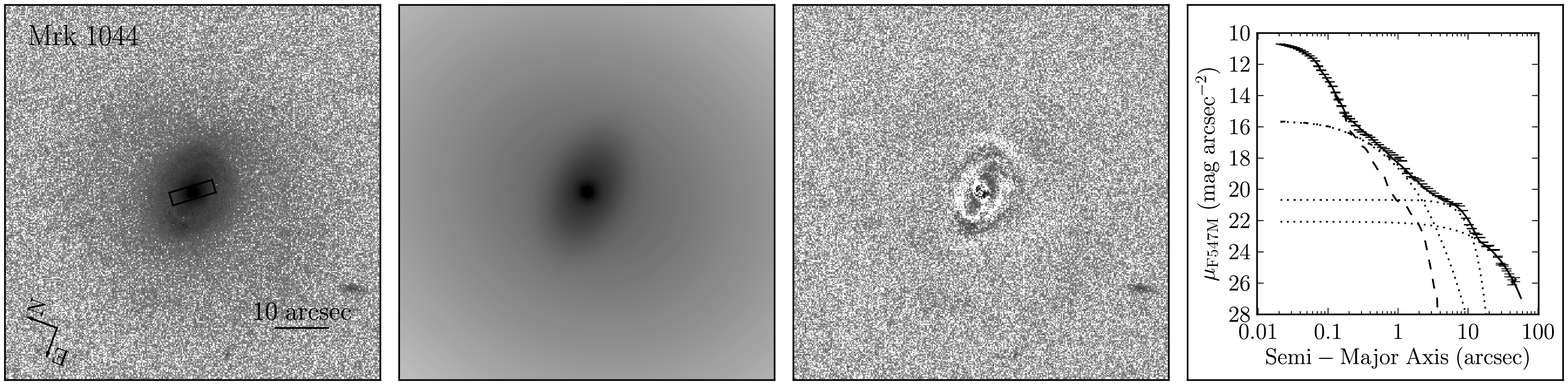} 
\includegraphics[angle=0,width=0.9\textwidth]{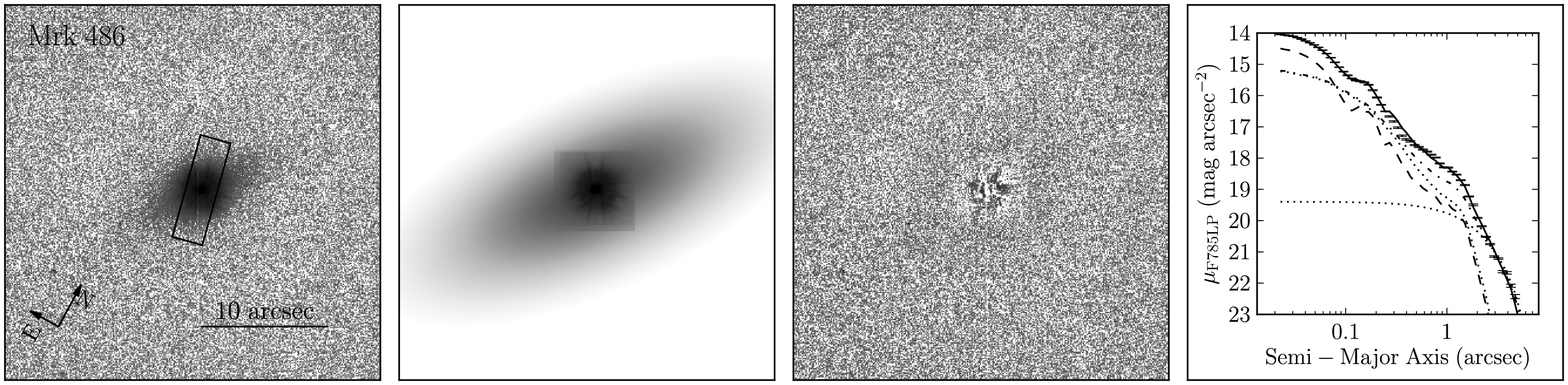} 
\includegraphics[angle=0,width=0.9\textwidth]{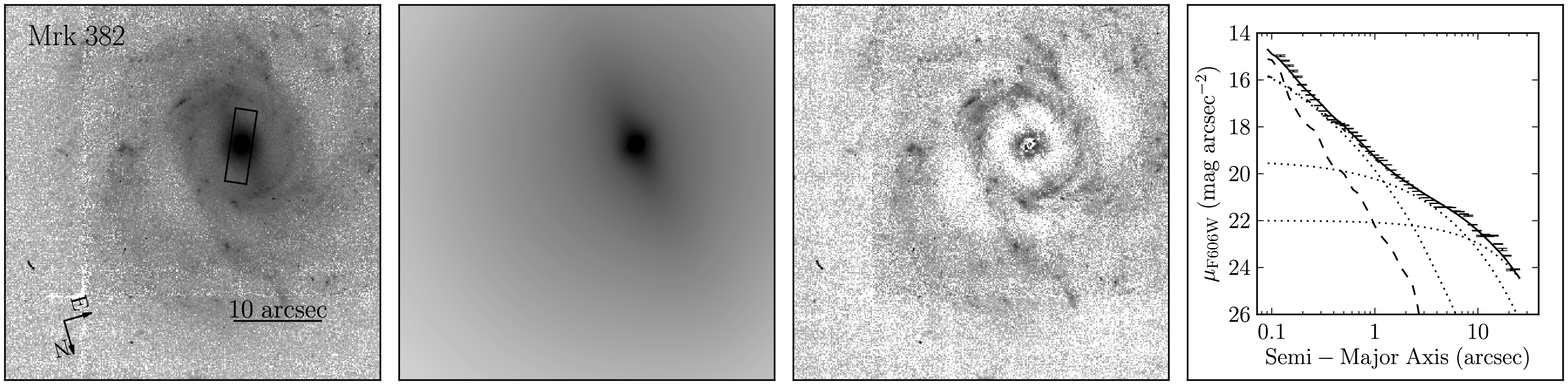}
\includegraphics[angle=0,width=0.9\textwidth]{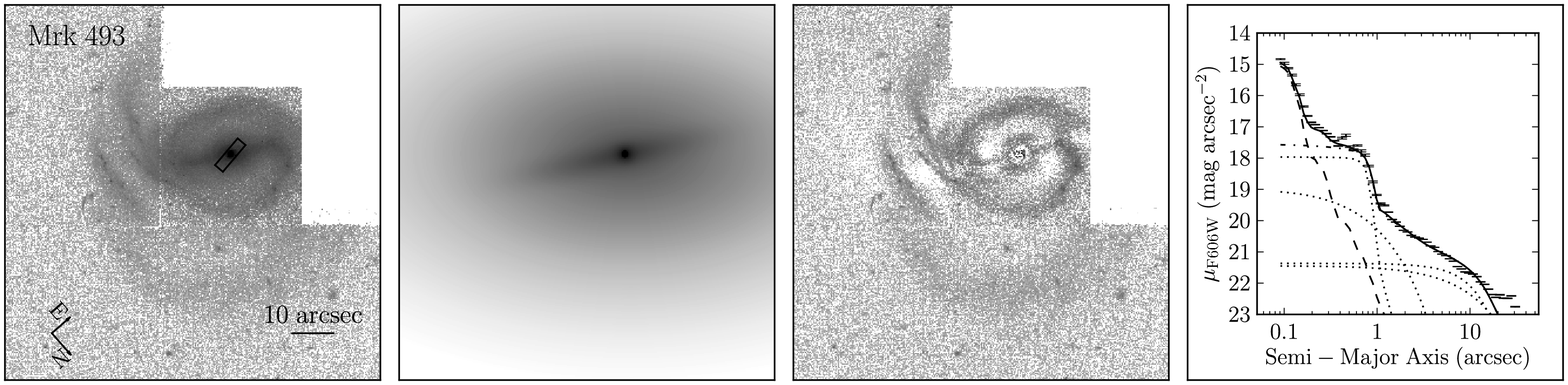} 
\includegraphics[angle=0,width=0.89\textwidth]{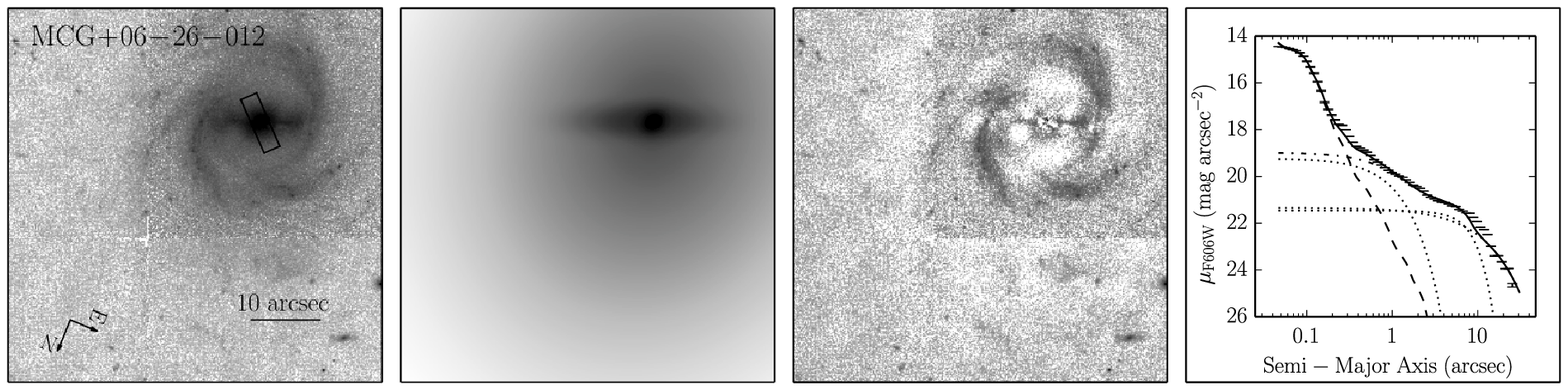} 
\end{center}
\caption{\footnotesize
Hubble Space Telescope images of Mrk 1044, IRAS F12397+3333, Mrk 486, Mrk 493 and
MCG +06-26-012. The
left panels show the original images and the small boxes illustrate the
spectroscopic aperture used to extract the spectrum.
The 2nd column shows model images, the 3rd one  the residuals obtained
after subtracting the fitted model. The
4th column shows one-dimensional surface profiles of the three galaxies.
Points with error bars are observed data, solid lines are the best-fit models,
dashed lines are PSFs, dash-dotted lines are host profiles and dotted lines 
are the components (S\'ersic profiles) used to model the host galaxy light.
}
\label{hst_image}
\end{figure*}

\section{Observations reduction and cross correlation analysis}
\subsection{Observations}
Detailed information of our RM campaign are given in paper I where we describe the observatory, 
the telescope, the spectrograph and the observing procedure in great detail. 
In short, targets were selected from the list of SEAMBH candidates in W13  based on their 
coordinates and the slope of their 2--10 keV continuum requiring $\Gammax\ge 2.0$. All targets 
are classified, spectroscopically, as NLS1s i.e., 1) FWHM(H$\beta) \lesssim 2000 \kms$; 
2) \oiii/H$\beta\lesssim 3$; 3) strong \feii\, emission lines.
Only radio-quiet sources are selected to avoid contamination by relativistic jet emission to
the optical continuum and, perhaps, emission lines\footnote{The radio-loud object 1H 0323+342 
was selected by chance to be included in our observations. We detected a time lag of the H$\beta$ 
line relative to the 5100\AA\ continuum and a couple of simultaneous $\gamma$-ray flares. These 
results will be reported separately (F. Wang, \etal\, 2014 in preparation).}. 
The campaign started in October 2012 and lasted until June 2013. We used the Lijiang 2.4-m 
telescope of Yunnan Observatory, in China. We obtained optical spectra of all 10 radio-quiet
selected sources almost every night with suitable weather conditions. Flux calibration is obtained 
through the use of a nearby comparison star that was observed in the same slit with the AGN. 
The variability of three of the sources (Mrk 335, Mrk 142 and IRAS F12397+3333), and their 
BH mass and accretion rate, was reported in paper I. Details of five additional sources, 
four of which were found to be SEAMBHs, are given in Table 1.
Two additional radio quiet sources, that do not show significant time lags, will
not be further discussed.

\begin{deluxetable*}{rrrccccccc}
\tablecolumns{10}
\tablewidth{0.95\textwidth}
\setlength{\tabcolsep}{4pt}
\tablecaption{Host galaxy decomposition}
\tabletypesize{\scriptsize}
\tablehead{
\colhead{Object}              & 
\colhead{Data set}            & 
\colhead{Observational setup} & 
\colhead{$m_{\rm st}^*$}      &
\colhead{$R_e$}               & 
\colhead{$n$}                 & 
\colhead{$b/a$}               & 
\colhead{P.A.}                &
\colhead{Note}                & 
\colhead{$\chi_{\nu}^2$} \\
& & & & ($''$) & & & (deg) & & 
}
\startdata
Mrk 486        & W0MT010 & WFPC, P6, F785LP &$16.69\pm0.01$  &      &       &      &       & PSF             & 1.278 \\
               &         &                  &$16.46\pm0.01$  & 0.11 & [2.0] &$0.66\pm0.01$ & $-45.73\pm1.04$& Bulge &       \\
               &         &                  & 16.75  & 2.41 & [1.0] & 0.41 &$-37.28\pm0.24$& Disk            &       \\
               &         &                  & 0.032  &      &       &      &       & Sky             &       \\
Mrk 382        & U2E62I01T & WFPC2, PC1, F606W  & 17.16  &      &       &      &        & PSF                 & 1.068 \\
               &           &                  & 16.77  & 0.51 &$3.34\pm0.03$ & 0.83 &$-64.65\pm0.42$ & Bulge   &       \\
               &           &                  &$16.25\pm0.02$ &$6.98\pm0.07$ &$1.70\pm0.02$& 0.41 &$7.38\pm0.07$  & Bar  &   \\
               &           &                  &$14.82\pm0.01$ &$18.5\pm0.11$ & [1.0] & 0.96 &$-43.38\pm3.52$ & Disk   &  \\
               &           &                  & 0.023  &      &       &      &        & Sky                 &       \\
MCG $+06$& U2E61L01T& WFPC2, PC1, F606W& 18.22  &      &       &      &        & PSF                 & 0.761 \\
               &           &                  & 18.43  & 0.07 &$0.35\pm0.01$ & 0.41 & $24.10\pm0.28$ & Add'l PSF           &       \\
               &           &                  & 18.09  & 1.02 & 0.81  & 0.82 &$-35.23\pm0.52$& Bulge               &       \\
               &           &                  & 17.05  &$6.48\pm0.01$ & 0.41  & 0.27 &$-67.93\pm0.04$ & Bar                 &       \\
               &           &                  & 14.69  &$15.4\pm0.02$ & [1.0] & 0.85 &$1.47\pm0.18$ & Disk                &       \\
               &           &                  & 0.019  &      &       &      &        & Sky                 &       \\
Mrk 493        & U2E62O01T & WFPC2, PC1, F606W& 17.34  &      &       &      &        & PSF                 & 1.191 \\
               &           &                  & 17.26  & 0.59 & 0.11  & 0.83 &$-38.13\pm0.24$ & N. Spiral/Ring &       \\
               &           &                  & 17.42  & 1.52 &$1.18\pm00.2$ & 0.85 &$49.24\pm0.62$ & Bulge               &       \\
               &           &                  & 16.23  &$11.4\pm0.01$ & 0.57  & 0.21 &$60.24\pm0.02$ & Bar                 &       \\
               &           &                  & 14.72  &$19.4\pm0.02$ & [1.0] & 0.57 &$48.41\pm0.05$ & Disk                &       \\
               &           &                  & 0.015  &      &       &      &        & Sky                 &       \\
Mrk 1044       & IBGU10  & WFC3, UVIS1, F547M & 15.35  &      &       &      &        & PSF                 & 1.160 \\
               &           &                  & 15.71  & 0.85 & 1.86  & 0.98 &$-87.24\pm2.07$ & Bulge               &       \\
               &           &                  & 15.38  & 6.33 & 0.41  & 0.65 &$ 87.63\pm0.08$ & Bar                 &       \\
               &           &                  & 14.70  &$21.2\pm0.11$ & [1.0] & 0.87 &$-3.48\pm0.50$ & Disk                &       \\
               &           &                  & 0.011  &      &       &      &        & Sky                 &       
\enddata
\tablecomments{The values in square brackets are fixed in the fitting procedure.
*$m_{\rm st}$ is the ST magnitude, an $f_{\lambda}$-based magnitude system,
$m_{\rm ST} = -2.5 \log_{10}(f_\lambda)-21.10$, for $f_\lambda$ in erg~s$^{-1}$cm$^{-2}$\AA$^{-1}$
(see Sirianni et al. 2005). The units of sky are electrons/s. Note that 
only errors that exceed a certain value are listed by GALFIT and cases without error bars mean an uncertainty
below this limit.
}
\end{deluxetable*}

\subsection{Host galaxy contamination}
All five targets were observed by the {\it Hubble Space Telescope} (HST) prior to our 
campaign. We use the HST images to remove the host galaxy contaminations as described 
in paper I. For objects with only one exposure, we use L. A. Cosmic (van Dokkum 2001) 
to remove cosmic rays in their images.  Unlike the three other sources, Mrk 486 has only 
WF/PC-1 exposure with a resolution which is too low to constrain the S\'ersic index of 
its bulge. Since a large fraction of NLS1s contain pseudo-bulges (Ryan \etal\,
2007; Orban de Xivry \etal\,  2011; Mathur \etal\, 2012), we fix the S\'ersic 
index of the bulge component  to 2.0. We also experimented with S\'ersic indexes of 
1.0 and 4.0 which changed the host flux by $\sim 10\%$, well within the uncertainty 
of this measurement.

The results of the fitting process are listed in Table 2 and  the images are shown in 
Figure 1. In most cases, the host contamination, given the slit size used in the observations 
(2.5$''$), does not amount to more than $\sim 25\%$ at $5100(1+z)$\AA.
The  uncertainty on the BH distance, $\Dbh$ (see below), due to this component is
$\Delta f_{\rm host}\lesssim 5\%$, where $f_{\rm host}$ is the fractional
contribution of the host at $5100(1+z)$\AA.

\subsection{Light curves and CC analysis}
The various panels of figure 2 show the light curves of the five new targets and Table 4 
presents the relevant measurements. 
We make use of the cross-correlation function (CCF) to measure the time lag 
of the  H$\beta$ line relative to the observed continuum. The interpolated cross-correlation 
function (ICCF; Gaskell \& Sparke 1986; Gaskell \& Peterson 1987) and the Z-transformed discrete
correlation function (ZDCF; Alexander 1997) methods were employed. 
The uncertainties of the time lags are determined using the ``flux randomization (FR)/random 
subset sampling (RSS)" method (Peterson \etal\, 1998a, 2004). More details are provided in 
paper I. In general, we prefer the use of the 5100\AA\ continuum for calculating the CCF. 
However, in one source, Mrk~382, we choose to use the better sampled $V$-band light curve despite 
the additional uncertainty due to the inclusion of several emission lines in this band.

In general, the variability we find in our hard X-ray selected NLS1 sample is 
different from what was found in previous monitoring of such objects where small amplitude 
variations in the optical bands were reported (Klimek \etal\, 2004).  The variations 
are consistent with several earlier suggestions that show the variability amplitude 
decreases with the Eddington ratio (e.g., Zuo \etal\, 2012; Ai \etal\, 2013). These properties
are important for the understanding of slim accretion disks and will be discussed in detail in 
future publications.

\begin{turnpage}
\begin{deluxetable*}{cccccccccccccccccccc}
\tablecolumns{16}
\tablewidth{0pt}
\setlength{\tabcolsep}{1.05pt}
\tablecaption{Continuum and H$\beta$ light curves}
\tabletypesize{\footnotesize}
\tablehead{
\multicolumn{3}{c}{Mrk 1044} & & \multicolumn{4}{c}{Mrk 382} & & \multicolumn{3}{c}{MCG $+06-26-012$} & & \multicolumn{3}{c}{Mrk 486} & & \multicolumn{3}{c}{Mrk 493} \\
\cline{1-3} \cline{5-8} \cline{10-12} \cline{14-16} \cline{18-20}
\colhead{JD} & \colhead{$F_{5100}$} & \colhead{$F_{\rm H\beta}$} & &
\colhead{JD} & \colhead{$V$} & \colhead{JD} & \colhead{$F_{\rm H\beta}$} & &
\colhead{JD} & \colhead{$F_{5100}$} & \colhead{$F_{\rm H\beta}$} & &
\colhead{JD} & \colhead{$F_{5100}$} & \colhead{$F_{\rm H\beta}$} & &
\colhead{JD} & \colhead{$F_{5100}$} & \colhead{$F_{\rm H\beta}$} }
\startdata
 29.3 & $ 5.35\pm 0.08$ & $ 3.82\pm 0.02$ & & 24.4 & $-0.074\pm 0.008$ & 24.4 & $ 0.41\pm 0.01$ & & 115.3 & $ 0.55\pm 0.01$ & $ 0.37\pm 0.01$ & & 179.4 & $ 3.54\pm 0.03$ & $ 3.45\pm 0.01$ & & 201.4 & $ 1.80\pm 0.01$ & $ 1.03\pm 0.01$ \\
 31.3 & $ 4.93\pm 0.03$ & $ 3.77\pm 0.02$ & & 25.4 & $-0.007\pm 0.009$ & 25.4 & $ 0.40\pm 0.01$ & & 117.3 & $ 0.55\pm 0.01$ & $ 0.39\pm 0.01$ & & 181.4 & $ 3.55\pm 0.08$ & $ 3.49\pm 0.02$ & & 202.4 & $ 1.76\pm 0.02$ & $ 1.00\pm 0.01$ \\
 36.2 & $ 4.89\pm 0.01$ & $ 3.76\pm 0.01$ & & 26.4 & $-0.021\pm 0.006$ & 26.4 & $ 0.40\pm 0.01$ & & 152.4 & $ 0.56\pm 0.01$ & $ 0.34\pm 0.01$ & & 185.4 & $ 3.48\pm 0.02$ & $ 3.41\pm 0.01$ & & 203.4 & $ 1.75\pm 0.01$ & $ 1.00\pm 0.01$ \\
 39.2 & $ 5.40\pm 0.03$ & $ 3.79\pm 0.01$ & & 27.4 & $-0.012\pm 0.005$ & 27.4 & $ 0.37\pm 0.01$ & & 153.4 & $ 0.56\pm 0.01$ & $ 0.33\pm 0.01$ & & 186.4 & $ 3.57\pm 0.05$ & $ 3.55\pm 0.03$ & & 204.4 & $ 1.74\pm 0.01$ & $ 1.01\pm 0.01$ \\
 41.2 & $ 5.53\pm 0.02$ & $ 3.92\pm 0.01$ & & 28.3 & $-0.022\pm 0.012$ & 28.3 & $ 0.39\pm 0.01$ & & 172.3 & $ 0.66\pm 0.01$ & $ 0.37\pm 0.01$ & & 190.4 & $ 3.46\pm 0.05$ & $ 3.56\pm 0.01$ & & 206.4 & $ 1.72\pm 0.01$ & $ 1.00\pm 0.01$ \\
\enddata
\tablecomments{\footnotesize
The full version of this table is also available in machine-readable form in the electronic version of the 
{\it Astrophysical Journal}. JD: Julian dates from 2456200; $F_{5100}$ and $F_{\rm H\beta}$ are fluxes at $(1+z)5100$\AA\, and H$\beta$
emission line in units of $10^{-15}\mathrm{~erg~s^{-1}~cm^{-2}~}$\AA$^{-1}$ and $10^{-13}\mathrm{~erg~s^{-1}~cm^{-2}}$, respectively.
$V$ is V-band instrumental magnitude. The systematic uncertainties of $F_{5100}$ and $F_{\rm H\beta}$ (see paper I) are
($\Delta F_{5100}$, $\Delta F_{\rm H\beta})= (0.163, 0.056), (0.018, 0.017)$, $(0.049, 0.052)$ and $(0.045, 0.025)$
for Mrk 1044, MCG 06, Mrk 486 and Mrk 493 respectively. $\Delta F_{\rm H\beta} = 0.016$ for Mrk 382. Since the space limitation, effective
numbers of JD and fluxes have been cut down in this Table, however they  
four effective numbers, all others have three, respectively, in the electronic version.  }
\end{deluxetable*}
\end{turnpage}

\begin{figure*}[t!]
\begin{center}
\includegraphics[angle=0,width=0.5\textwidth]{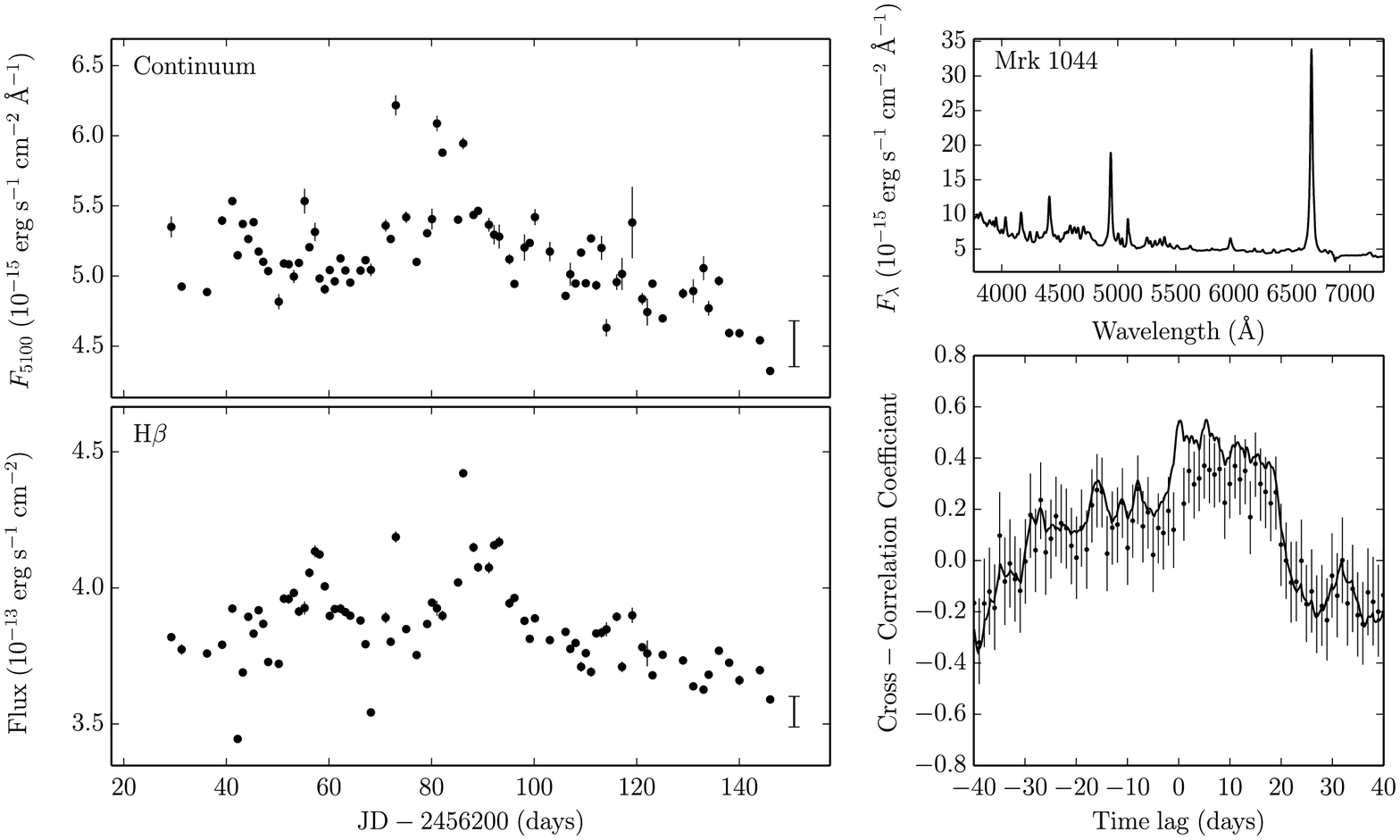}
\includegraphics[angle=0,width=0.5\textwidth]{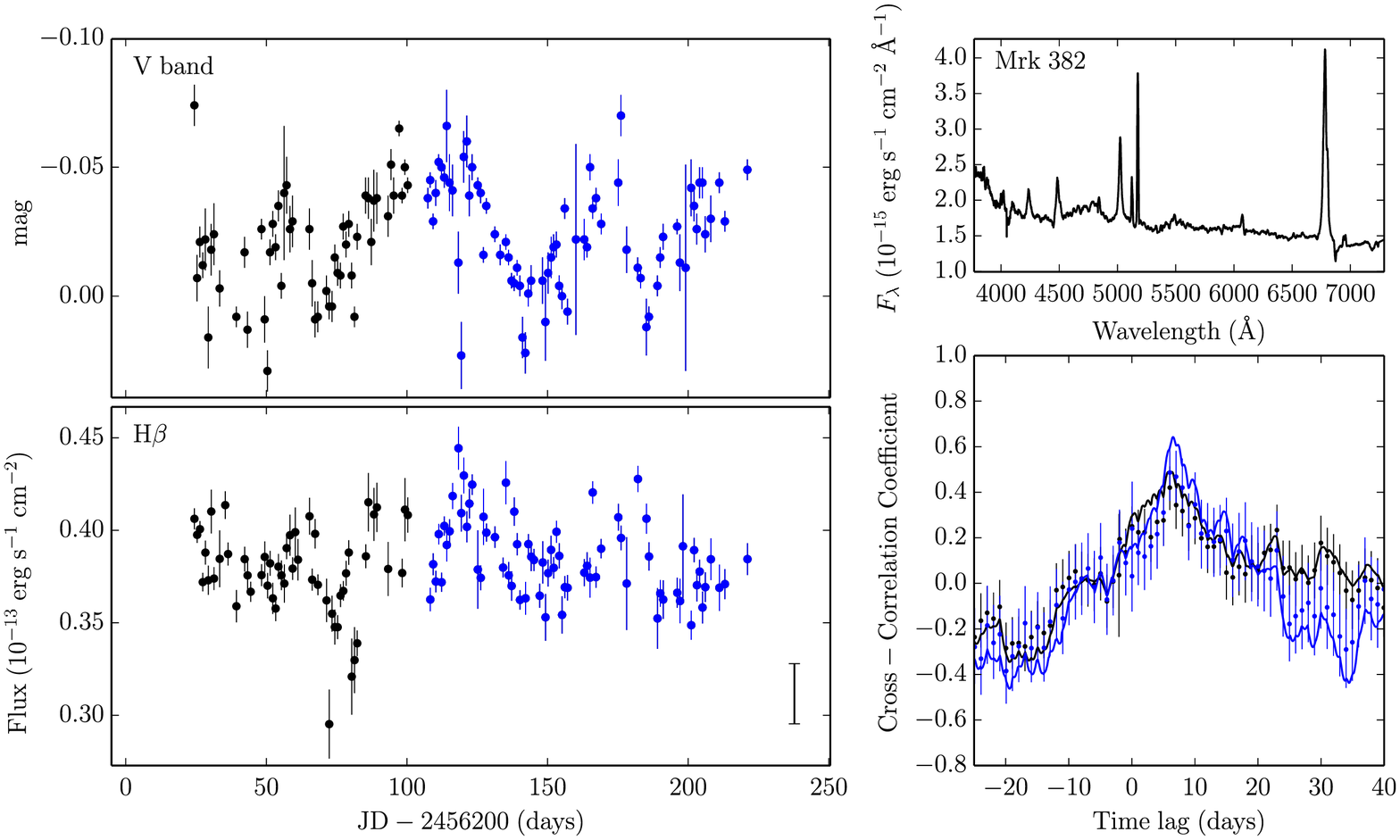}
\includegraphics[angle=0,width=0.5\textwidth]{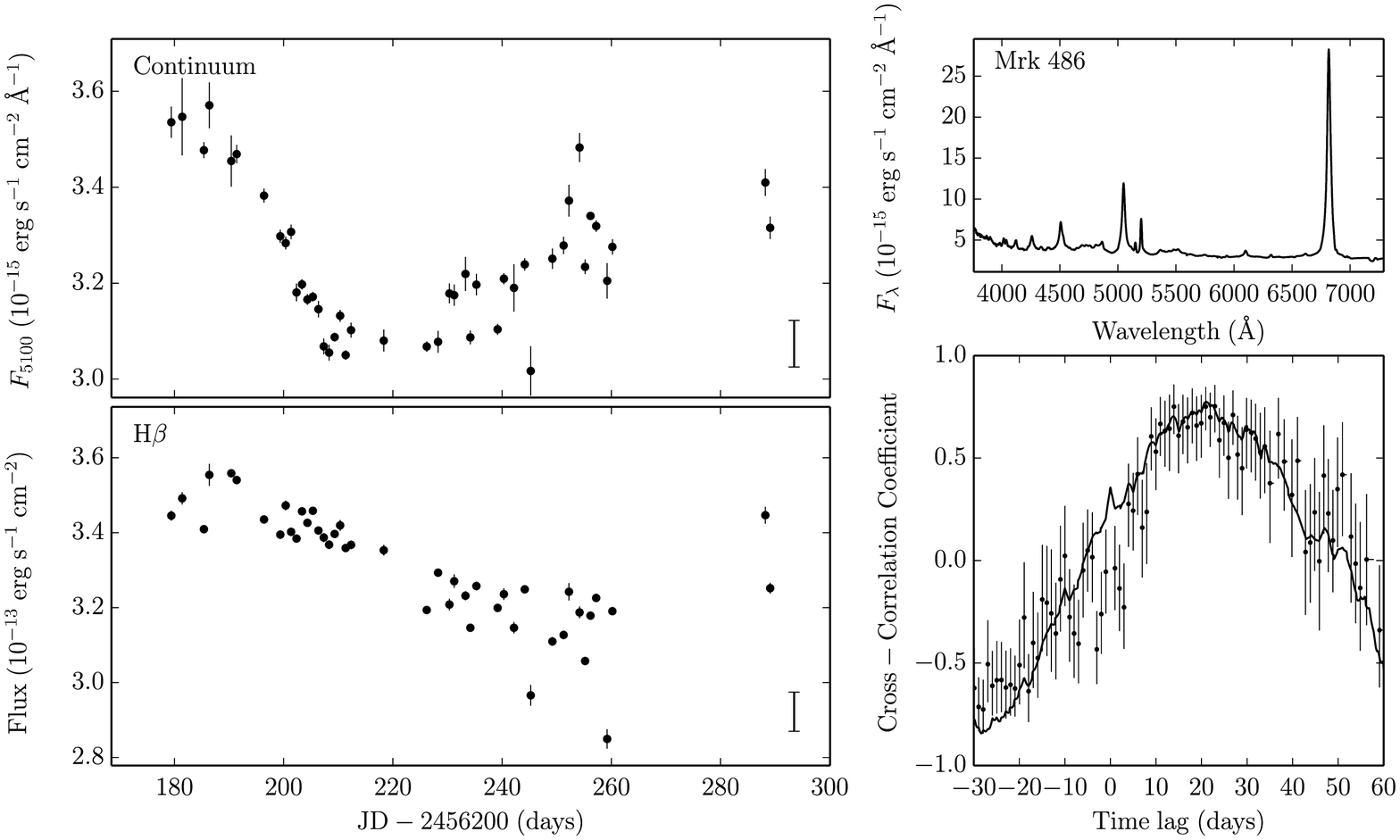}
\includegraphics[angle=0,width=0.5\textwidth]{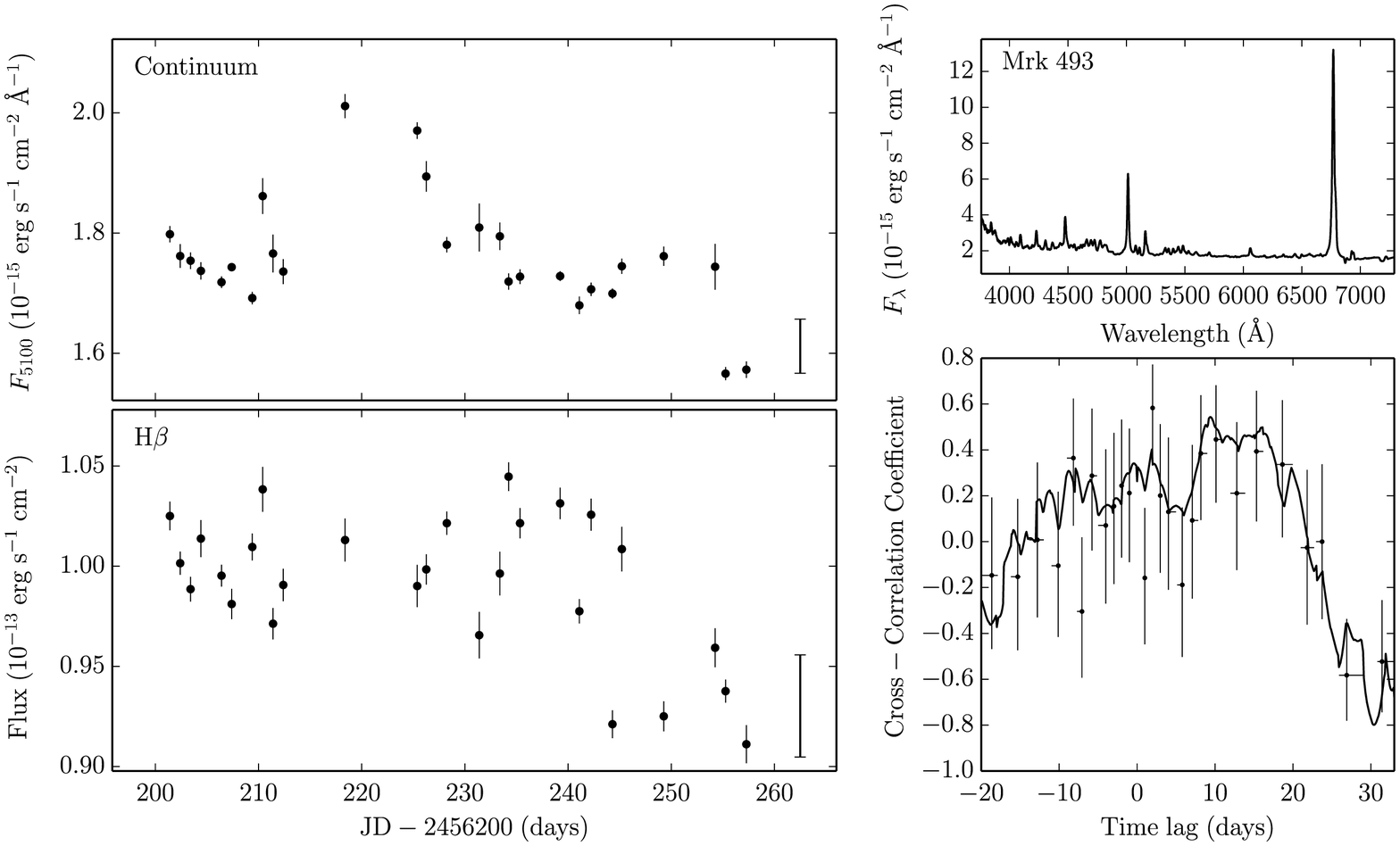}
\includegraphics[angle=0,width=0.5\textwidth]{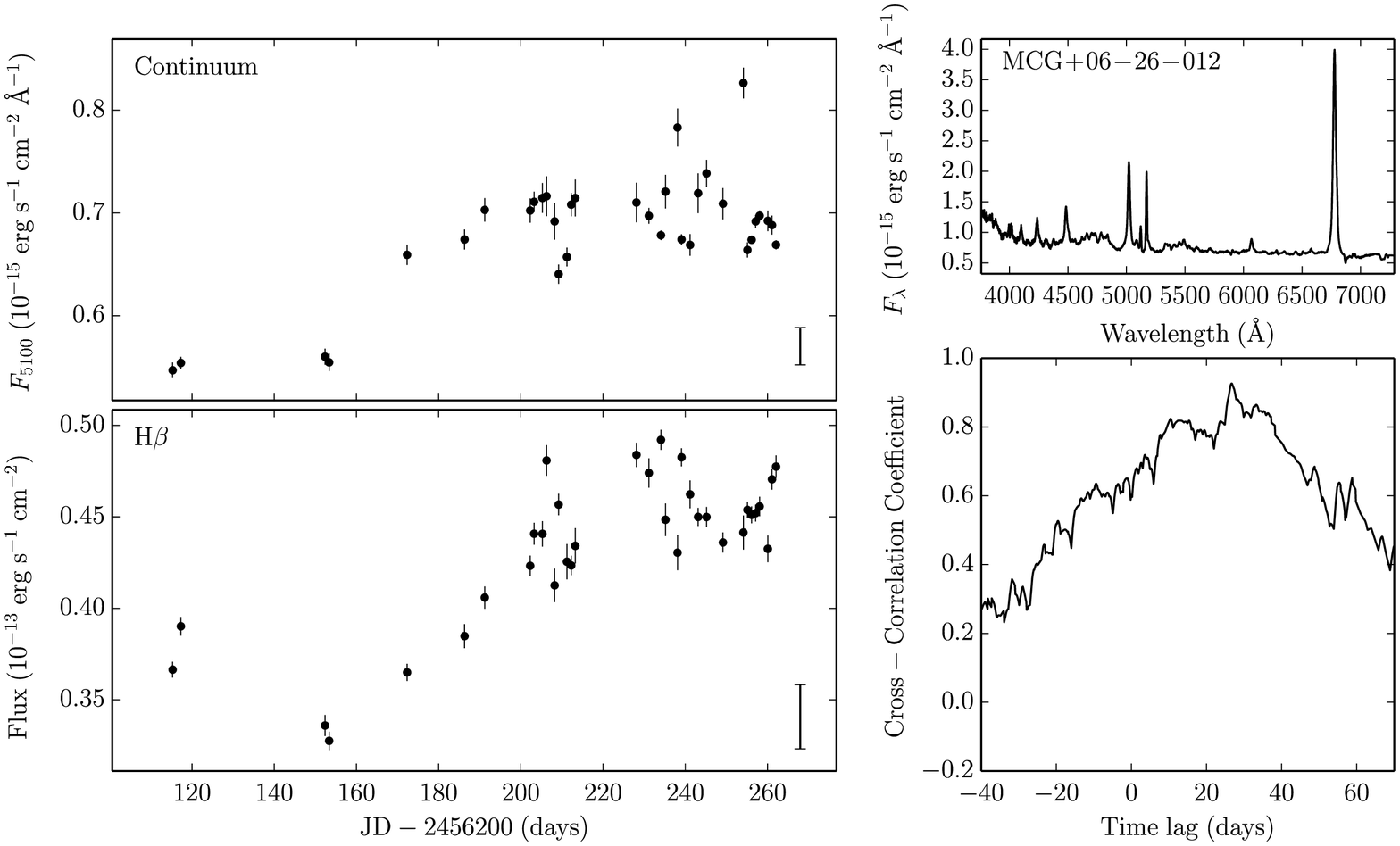} 
\end{center}
\vspace{-0.63cm}
\caption{\footnotesize
Light curves and CCF analysis.
The {\it left} panels show light curves of H$\beta$ and $F_{5100}$ ($V$-band in Mrk 382).
The {\it right} panels show mean spectra and cross-correlation functions. For Mrk 382,
we divided the light curves into two parts shown by blue and black points. The black CCFs
are obtained from the entire light curves while the blue one from the blue light curves.
The parts of the light curves shown by blue points result in a more significant time lag 
which is the one listed in Table 1. 
}
\end{figure*}

\subsection{Notes on individual sources}
{\it Mrk 1044}: {\it ROSAT} observations show evidence for soft X-ray variability
of this source (Boller \etal\, 1996). Our CCF analysis shows a peak at $\sim 5$ days with a
maximum cross-correlation coefficient of $r_{\rm max}\approx 0.55$. The measured $\bhm$
combined with the conservative assumption on the spin give
$\mdotmin\sim 2.73$, the largest in our sample.

{\it Mrk 382}: This object shows the strongest host contamination in our sample.
Because of this, the 5100\AA\ flux calibration is uncertain and we chose to
use the $V$-band light curve, with their improved precision, in the CCF
analysis (the only source in our sample). Using the entire
light curve we obtain a peak correlation of $r_{\rm max}=0.49$ and a time lag
of $\tau = 5.8^{+1.6}_{-1.5}$ days. Since the first part of the observations of this
object suffers from bad-weather and problems with continuum subtraction
from the H$\beta$ line, this resulted in a noisier light curve in
the first part of the campaign. Thus, we also analyzed only the second
half of the light curve, with its improved sampling (blue points in
Figure 2). This gives a stronger correlation with
$r_{\rm max}=0.65$ corresponding to a time lag of $\tau = 6.6^{+1.1}_{-0.7}$ 
days in the rest frame of the source. In the following analysis we use this 
time lag.

{\it Mrk 486}: The source shows significant flux and SED variations in the X-ray
band (Ballo \etal\, 2008). The gap in the data between JD260-290 (see Figure 2) 
is due to bad weather. The H$\beta$ light curve shows a monotonic decreases with 
time and much of the signal in the CCF is due to the flux increase at the end of 
the campaign, on July 14 and 15, 2013. The observations on those night were taken 
under very good conditions and we have no reason to suspect these measurements.

{\it Mrk 493}: We obtained  only 27 observations from April to June 2013. This makes
the quality of the CCF  poorer than in other objects and results in a larger uncertainty
on the time lag which is consistent with zero (although the peak of the CCF is very clear 
with $r_{\rm max}\sim 0.54$). We report on the various results obtained for this source
but do not include it in the distance analysis.

{\it MCG +06-26-012:} This NLS1 is situated in a Sb galaxy (see {\it HST} image in Figure 1). 
We find FWHM(H$\beta$)=1685 $\kms$ (see also Grupe \etal\, 1999; Veron-Cetty \etal\, 2001). 
The CCF shows a statistically significant time lag of $\taublr= 23.3_{-5.8}^{+7.5}$ 
days with $r_{\rm max}\approx 0.9$. From Eqn. (\ref{eq:virial}), we get 
$\bhm=(8.3_{-2.3}^{+2.9})\times 10^6\sunm$. The host-galaxy subtracted luminosity is 
$L_{5100}= (0.47\pm 0.10)\times 10^{43}\ergs$. Using our method to determine accretion 
rate [Eqn. (\ref{eq:dotm_min}) below] we get $\dot{m}_{\rm min}=0.02$, 
indicating that this object is not a SEAMBH. 

Finally we comment on the general method of measuring line intensities in this and earlier 
RM experiments. The H$\beta$ light curves reported here and in paper I were obtained
by simple integration over the line profile using a locally determined continuum.
These are not necessarily the most accurate light curves and spectral fitting methods that 
include other components, such as the host galaxy SED and the \feii\ lines (e.g., Barth et 
al. 2013) can lead to significant improvements and reduced uncertainties. In some rare cases, 
this can also be used to recover points in the light curve that were discarded due to poor 
weather conditions that prevented us from using the local comparison star as our flux calibrator. 
We are working on the improvement of such methods and will report the
results in a future publication (Hu et al. in preparation).

\section{Super-Eddington accreting massive black holes}
\subsection{Black hole mass measurements}
Measuring BH mass through RM is a well established method.
It is based on the idea that the velocities of the clouds emitting the
broad emission lines are virialized in the BH gravitational
potential and the emitted line intensities echo the variable ionizing continuum. The
time-lag, $\taublr$,  reflects the size and geometry of the variable part of the BLR
and the emissivity weighted radius is given by
$\rblr=c\taublr$,
where $c$ is the speed of light. RM experiments provide the necessary measurements of 
$\rblr$ which, combined with the virial assumption,  result in a simple expression for 
the BH mass,
\begin{equation}
\bhm=\fblr \frac{V_{\rm BLR}^2\rblr}{G},
\label{eq:virial}
\end{equation}
where $G$ is the gravitational constant, $V_{\rm BLR}$ is a measure of the gas velocity, 
and $\fblr$ is a constant which combines all the unknowns about the geometry and kinematics 
of the gas in the BLR. The best value of $\fblr$ is obtained by using BH mass estimated 
determined by the $\bhm-\sigma_*$ relationship in AGN hosts where stellar absorption features 
can be observed (Woo et al. 2013).

We have used our observations to measure the mass of the BHs in our sample through 
Eqn. (\ref{eq:virial}).
In principal there are four possible choices to define $V_{\rm BLR}$ (e.g., Collin \etal\, 2006):
1) the full-width-at-half-maximum (FWHM) of the variable component of the line obtained from 
the RMS spectrum (Peterson \etal\, 1998b; Wandel \etal\, 1999), 2) the FWHM of the line 
obtained from the mean spectrum (e.g., Kaspi \etal\, 2000), 3) the line dispersion (so called 
$\sigma_{\rm line}$) obtained from the rms spectrum (e.g., Fromerth \& Melia 2000; Peterson \etal\, 
2004; Collin \etal\, 2006; Denney \etal\, 2013) and 4) the line dispersion obtained from the mean 
spectrum (e.g., Collin \etal\, 2006; Bian \etal\, 2008). For example, the recent Woo et al. (2013) 
work suggests that for $\sigma_{\rm line}$ measured from the rms spectrum, $\fblr\simeq 5.3$. Despite 
many years of study, there is no empirical evidence, or a realistic model,  to show that any of the 
methods is preferred over the others. 

Following paper I, we have chosen to use the FWHM(H$\beta$) from the mean spectrum as
our choice of $V_{\rm BLR}$. 
Applying this to the Woo et al. (2013) sample of 25 AGNs with measured $\sigma_*$ gives 
$\fblr =  1.00\pm 0.25$ (Woo, private communication) with no dependence on BH mass or accretion 
rate. The scatter in the derived $\bhm$ obtained in this way is very similar to the scatter observed 
by assuming $V_{\rm BLR}=\sigma_{\rm line}$ with its corresponding $\fblr$. Our measurements of 
$\bhm$ based on this method are listed in Table 4 and more references regarding this choice are 
provided in paper I.

\subsection{Identifying SEAMBHs by their accretion rates}
The SEAMBHs discussed in this work belong to the category of slim accretion disks.
They can be recognized by their $\mdot$ that approaches and even exceeds unity. 
However, the uncertainties on present slim disk models, and the uncertainties on the  
observations, are too large to identify a specific value of $\mdot$ beyond which photon 
trapping is significant. Various models (Laor \& Netzer 1989; Beloborodov 1998; Sadowski 
\etal\, 2011) suggests that this occurs at $\mdot \simeq 0.1-0.3$ but even this
range is uncertain.

Returning to SEAMBHs with higher accretion rate, we note that the value of $\mdot$ which 
is required to identify such objects can be determined from the measured $\bhm$ and the 
global SED of the disk. However, we do not have access to the entire SED because of the 
Galactic and intergalactic absorption. We can only estimate the mass accretion rates, 
$\Mdot$, from the observed optical continuum and the results of the 
existing slim disk models that suggest this to be a good approximation since the wavelengths 
used correspond to the parts of the disk that are not affected by photon trapping. The 
uncertainty on BH spin, and $\mathdotM$ and $\eta$ still remains.

To overcome the uncertainty related to the BH spin, we adopted a conservative approach that
uses the smallest possible $\mdot$ based on the smallest possible $\eta$. 
For a maximally rotating pro-grade disk ($a=0.998$), the radiative efficiency reaches its
maximum of $\eta_{\rm max}=0.32$. From Eqn.~(\ref{eq:SS_2}), the maximum accretion rate assuming
$\cos i=0.75$ is $\dot{m}_{\rm max}=9.9~l_{44}^{3/2}m_7^{-2}$. If $\dot{m}_{\rm max}\le 0.1$, 
this object has too low accretion rate to maintain a slim disk.
In the same way we define a minimum accretion rate, $\dot{m}_{\rm min}$ by using the
lowest possible $\eta_{\rm min}=0.038$ corresponding to a retrograde spinning BH with $a=-1$,
\begin{equation}
\dot{m}_{\rm min}=1.2~l_{44}^{3/2}m_7^{-2} \, .
\label{eq:dotm_min}
\end{equation}
If  $\dot{m}_{\rm min}\ge 0.1$, the disk must be slim, and the object is considered to be 
a SEAMBH. Obviously there are
objects with $\dot{m}_{\rm min} < \dotm < \dot{m}_{\rm max}$ which we do not consider
SEAMBHs by our very conservative criterion, that may be powered by slim accretion disks
if $\eta$ is large enough. We can introduce an even more conservative criterion  
by combining $\mdot_{\rm min}$ and $\mdot_{\rm max}$. Since
$\mdot_{\rm min}=\left(\eta_{\rm min}/\eta_{\rm max}\right)\mdot_{\rm max}\approx 0.12\mdot_{\rm max}$,
we get $\mdot_{\rm min}\ge 0.12$ for SEAMBHs with $\mdot_{\rm max}\ge1$. In the present paper
we use $\mdot_{\rm min}\ge 0.1$ as the only criterion.

\begin{figure}[t!]
\begin{center}
\includegraphics[angle=0,width=0.4\textwidth]{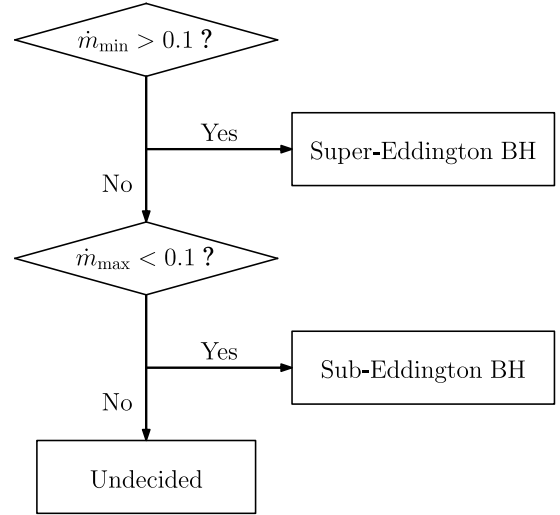} 
\end{center}
\caption{\footnotesize
Flow chart used to identify SEAMBHs. The undecided category is for sources
with not enough information on $\dotm$ to decide whether they are powered buy thin or
slim disks.}
\label{dotm_criterion}
\end{figure}
\vglue 0.5cm
\renewcommand{\arraystretch}{1.5}
\begin{deluxetable*}{lccccccrc}
\tablecolumns{8}
\setlength{\tabcolsep}{3pt}
\tablewidth{0pc}
\tablecaption{SEAMBHs: masses and accretion rates}
\tabletypesize{\scriptsize}
\tablehead{
\colhead{Objects}                  &
\colhead{$\taublr$}                &
\colhead{FWHM}                     &
\colhead{$\bhm$}                   &
\colhead{$\mdotmin$}               &
\colhead{$F_{\lambda}[(1+z)5100{\rm \AA}]$} &
\colhead{$E$(B$-$V)}               &
\colhead{$\Dl$}                    &
\colhead{$\Dbh$}                    \\ \cline{2-9}
\colhead{}                         &
\colhead{(days)}                   &
\colhead{($\kms$)}                 &
\colhead{($10^6\sunm$)}            &
\colhead{}                         &
\colhead{($10^{-15}\ergs{\rm cm^{-2}}\AA^{-1}$)}&
\colhead{}                         &
\colhead{(Mpc)}                    &
\colhead{(Mpc)}
}
\startdata
\multicolumn{9}{c}{SEAMBHs identified by the Shangri-La campaign}\\ \hline
Mrk 335     & $ 10.6_{- 2.9}^{+ 1.7}$ & $1997\pm265$ & $  8.3_{- 3.2}^{+ 2.6}$ &$0.60_{-0.29}^{+1.13}$& $ 5.20\pm0.37$ &0.030& 117.6 & $ 85.9_{-26.3}^{+21.5}$ \\
Mrk 1044    & $  4.8_{- 3.7}^{+ 7.4}$ & $1211\pm 48$ & $  1.4_{- 1.1}^{+ 2.1}$ &$2.73_{-2.38}^{+60.63}$& $ 3.28\pm0.37$ &0.031&  74.5 & $ 31.4_{-20.8}^{+32.2}$ \\
Mrk 382     & $  6.6_{- 0.7}^{+ 1.1}$ & $1588\pm330$ & $  3.3_{- 1.4}^{+ 1.5}$ &$0.54_{-0.34}^{+1.55}$& $ 0.78\pm0.13$ &0.043& 154.5 & $114.2_{-42.4}^{+47.4}$ \\
Mrk 142     & $  6.4_{- 2.2}^{+ 0.8}$ & $1647\pm 69$ & $  3.4_{- 1.2}^{+ 0.5}$ &$2.25_{-0.84}^{+4.11}$& $ 1.27\pm0.15$ &0.015& 207.9 & $ 95.5_{-28.6}^{+16.4}$ \\
IRAS F12397& $ 11.4_{- 1.9}^{+ 2.9}$ & $1835\pm473$ & $  7.5_{- 4.1}^{+ 4.3}$  &$0.51_{-0.33}^{+2.27}$& $ 1.44\pm0.14$ &0.017& 201.0 & $154.5_{-68.4}^{+67.8}$ \\
Mrk 486     & $ 20.0_{- 3.2}^{+ 8.7}$ & $1926\pm157$ & $ 14.5_{- 3.3}^{+ 6.7}$ &$0.19_{-0.11}^{+0.17}$& $ 2.34\pm0.17$ &0.012& 179.3 & $192.8_{-37.3}^{+67.9}$ \\
Mrk 493     & $ 12.2_{-16.7}^{+ 3.5}$ & $ 784\pm 11$ & $  1.5_{- 1.5}^{+ 0.4}$ &$>1.23$               & $ 0.94\pm0.13$ &0.022& 143.5 & $ 62.0_{-62.0}^{+17.4}$                  \\ \hline
\multicolumn{9}{c}{SEAMBHs identified from sources mapped by previous campaigns}\\ \hline
Mrk 110     & $ 24.3_{- 8.3}^{+ 5.5}$ & $1543\pm  5^a$ & $ 11.3_{- 3.9}^{+ 2.6}$ &$0.30_{-0.11}^{+0.43}$& $ 2.79\pm0.09$ &0.011& 162.1 & $ 149.2_{-39.0}^{+25.1}$ \\
Mrk 202     & $  3.0_{- 1.1}^{+ 1.7}$ & $1471\pm 18^b$ & $  1.3_{- 0.5}^{+ 0.7}$ &$0.17_{-0.13}^{+0.46}$& $ 0.30\pm0.11$ &0.018&  95.5 & $ 100.8_{-36.1}^{+62.3}$ \\
NGC 4051    & $  1.9_{- 0.5}^{+ 0.5}$ & $1453\pm  3^a$ & $  0.8_{- 0.2}^{+ 0.2}$ &$0.17_{-0.10}^{+0.24}$& $ 4.93\pm1.00$ &0.011&  17.1 & $ 17.9_{- 4.6}^{+  5.7}$ \\
NGC 7469    & $ 24.3_{- 4.0}^{+ 4.0}$ & $1722\pm 30^a$ & $ 14.1_{- 2.4}^{+ 2.4}$ &$0.17_{-0.06}^{+0.11}$& $10.80\pm1.00$ &0.061&  73.9 & $ 82.4_{-13.0}^{+13.9}$ \\
PG 0026+129 & $111.0_{-28.3}^{+24.1}$ & $2544\pm 56^a$ & $140.3_{-36.3}^{+31.1}$ &$0.17_{-0.06}^{+0.15}$& $ 2.31\pm0.07$ &0.063& 700.4 & $829.9_{-163.9}^{+136.1}$ \\
PG 0844+349 & $ 32.3_{-13.4}^{+13.7}$ & $2694\pm 58^a$ & $ 45.8_{-19.1}^{+19.5}$ &$0.12_{-0.07}^{+0.32}$& $ 2.57\pm0.38$ &0.033& 300.0 & $390.3_{-139.9}^{+151.8}$ \\
PG 1211+143 & $ 93.8_{-42.1}^{+25.6}$ & $2012\pm 37^a$ & $ 74.2_{-33.4}^{+20.4}$ &$0.26_{-0.14}^{+0.86}$& $ 5.06\pm0.92$ &0.030& 383.6 & $386.8_{-151.4}^{+119.0}$ \\
PG 1700+518 & $251.8_{-38.8}^{+45.9}$ & $2252\pm 85^a$ & $249.5_{-42.8}^{+49.2}$ &$0.45_{-0.14}^{+0.22}$& $ 1.86\pm0.03$ &0.030&1565.9 & $1358.1_{-171.9}^{+188.1}$ \\
PG 2130+099 & $ 31.0_{- 4.0}^{+ 4.0}$ & $2450\pm188^c$ & $ 36.3_{- 7.3}^{+ 7.3}$ &$0.18_{-0.06}^{+0.11}$& $ 2.53\pm0.09$ &0.039& 295.0 & $332.8_{-52.0}^{+50.9}$
\enddata
\tablecomments{\footnotesize
All measurements of Mrk 335, Mrk 142 and IRAS F12397 are from paper I. The other objects of the 
campaign are from this work. $a$: FWHM is from Collin \etal\, (2006); $b$: FWHM from Bentz 
\etal\. (2009a); $c$: FWHM from Grier \etal\, (2012). $F_{\lambda}[(1+z)5100{\rm \AA}]$ is 
obtained from the mean spectrum after the subtraction of the host galaxy contribution.
For the previous campaigns, we use the values of $F_{\lambda}[(1+z)5100{\rm \AA}]$
and $\taublr$ corrected by Bentz \etal\, (2009ab) and Bentz \etal\, (2013). $E(B-V)$ is the Galactic
extinction using the maps in Schlafly \etal\, (2011). All listed
values of $\dl$ are obtained from the redshift using standard cosmology (Ade \etal\, 2013)
except for NGC~4051 where the redshift is very small and a more reliable value of
$\dl=17.1$ Mpc is obtained from the Tully-Fisher relation (Bentz \etal\, 2013). 
The error on $\dl$ is assumed to be 2\% in all sources (see text). The errors
on $\dbh$ are from the calculations of $\Delta_{\rm obs}$ (the combination of the errors on $F_{5100}$, 
FWHM and $\taublr$). Note that the mass and distance calculated for Mrk 493 are consistent with zero 
and hence this source is not used in the distance analysis.
}
\end{deluxetable*}

A flow chart diagram for identifying SEAMBHs is given in Figure 3. Applying Eqn.~(\ref{eq:dotm_min})  
to our sample we find 7 targets (the three from paper I and four from this work) to have slim 
accretion disks. As explained earlier, we decided not to use Mrk\,493 because of the large 
uncertainty on the time lag and hence the BH mass. All together, six of the new objects are suitable 
for the distance analysis. 
It is  interesting to note that the fraction of SEAMBHs from our targets reaches 
$\sim 70\%$ suggesting that the $\Gamma_{2-10}$-based selection is very efficient for identifying 
SEAMBH candidates.

We used the same method to search for SEAMBHs among the entire sample ($\sim 40$) of radio-quiet 
AGNs with previous RM-based BH mass measurements (Peterson \etal\, 1998a; Kaspi \etal\, 2000, 2005; 
Bentz \etal\, 2013). We found 9 objects with $\mdot_{\rm min}>0.1$. Thus the total number of newly
identified SEAMBHs with directly measured BH mass and $\dotm_{\rm min}>0.1$, is 15 under the most 
conservative assumption about the BH spin. We also found 21 objects with $\dotm_{\rm max}<0.1$, 
which must be powered by thin disks. The rest of the sources have 
$\mdot_{\rm min} < \mdot < \mdot_{\rm max}$. 
Obviously the real number of objects containing slim disks could be larger,
but because of our conservative estimate of $\dot{m}$, we have no way to prove
it. The uncertainty on $\dotm_{\rm min}$ is obtained from the uncertainty on $\dot{M}$ that 
includes the measured flux, the BH mass and the inclination to the line-of-sight. The uncertainty 
due to inclination is discussed below.

\begin{figure*}[t!]
\begin{center}
\includegraphics[angle=0,width=0.31\textwidth]{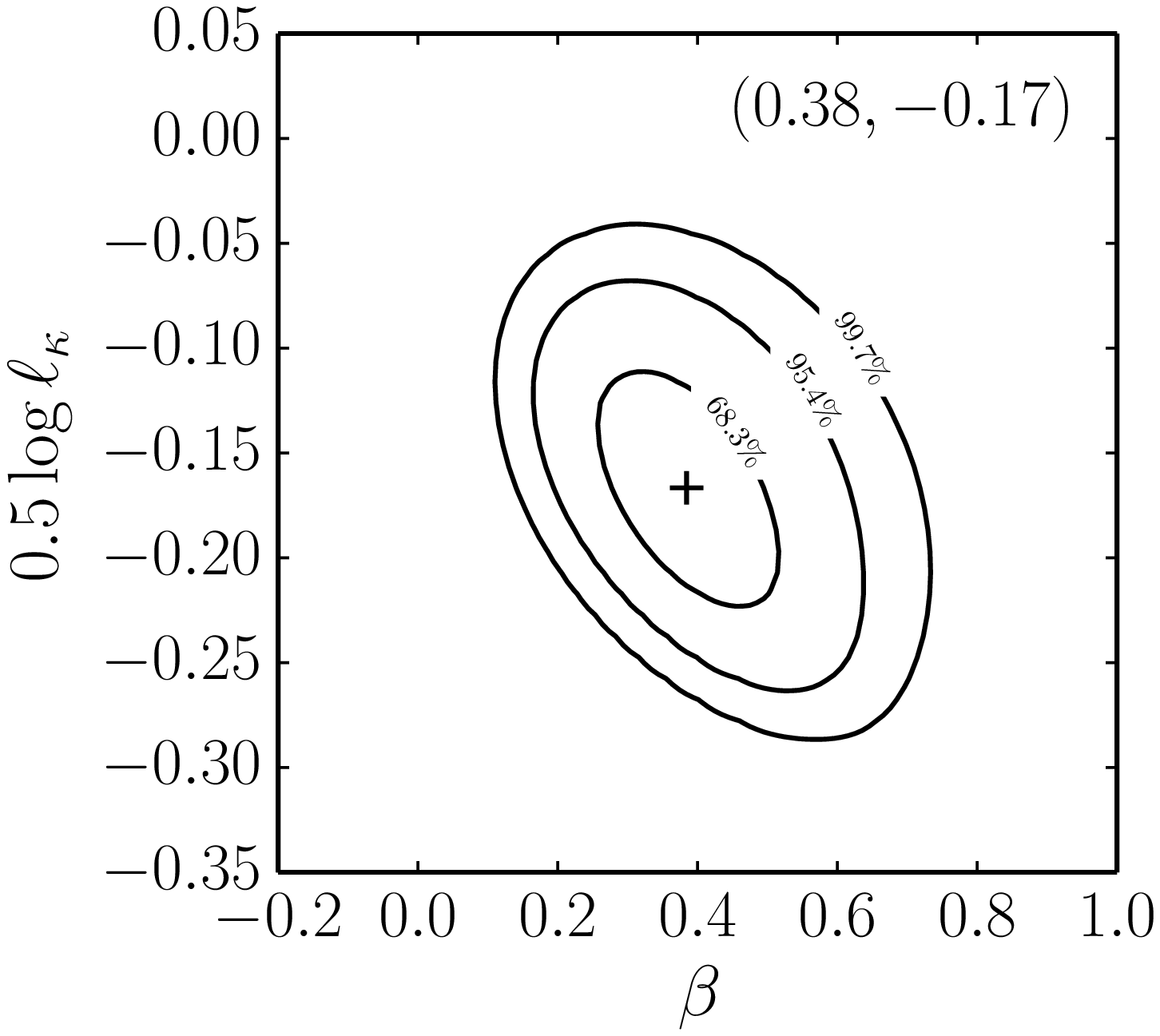}
\includegraphics[angle=0,width=0.31\textwidth]{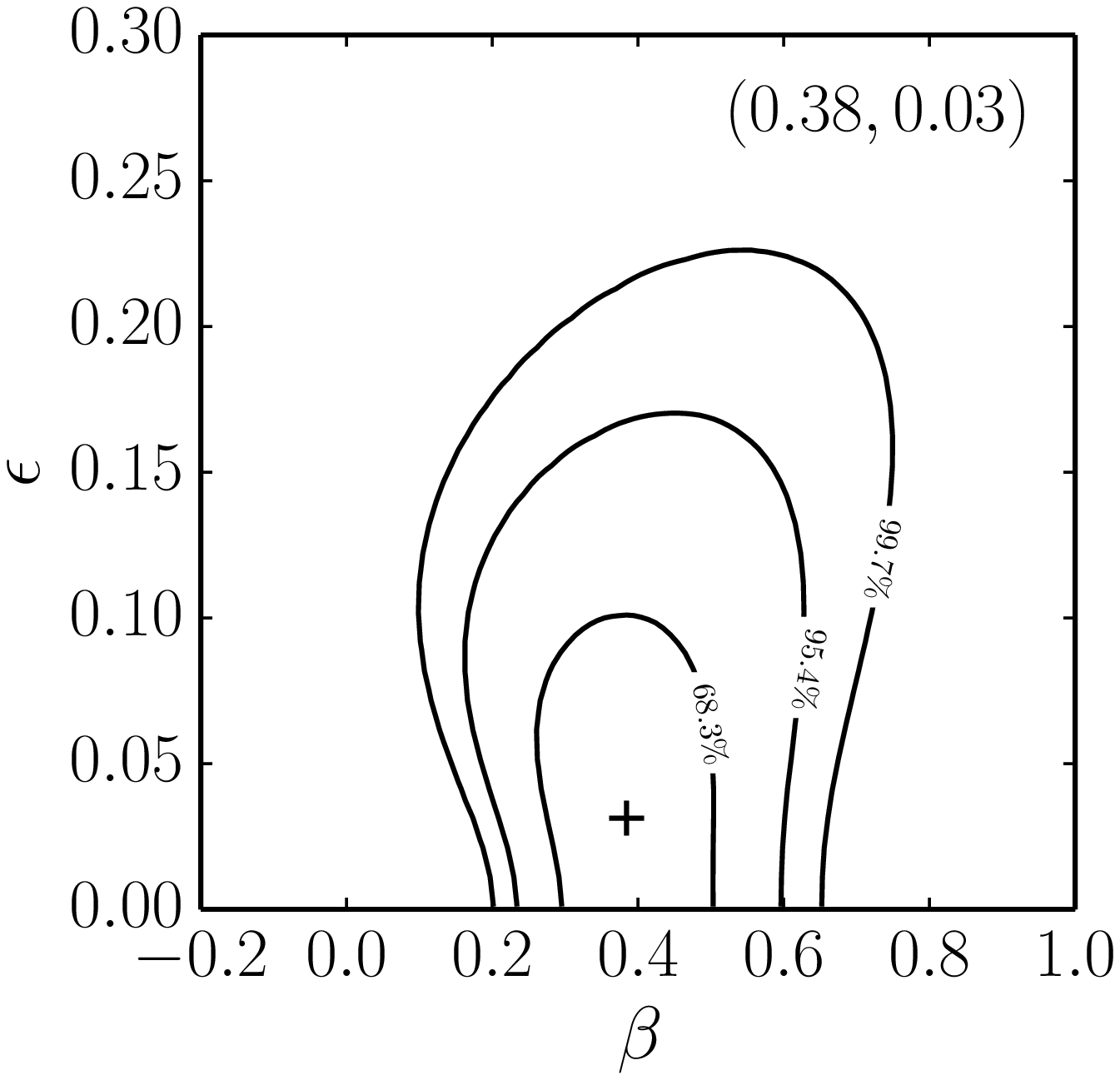}
\includegraphics[angle=0,width=0.31\textwidth]{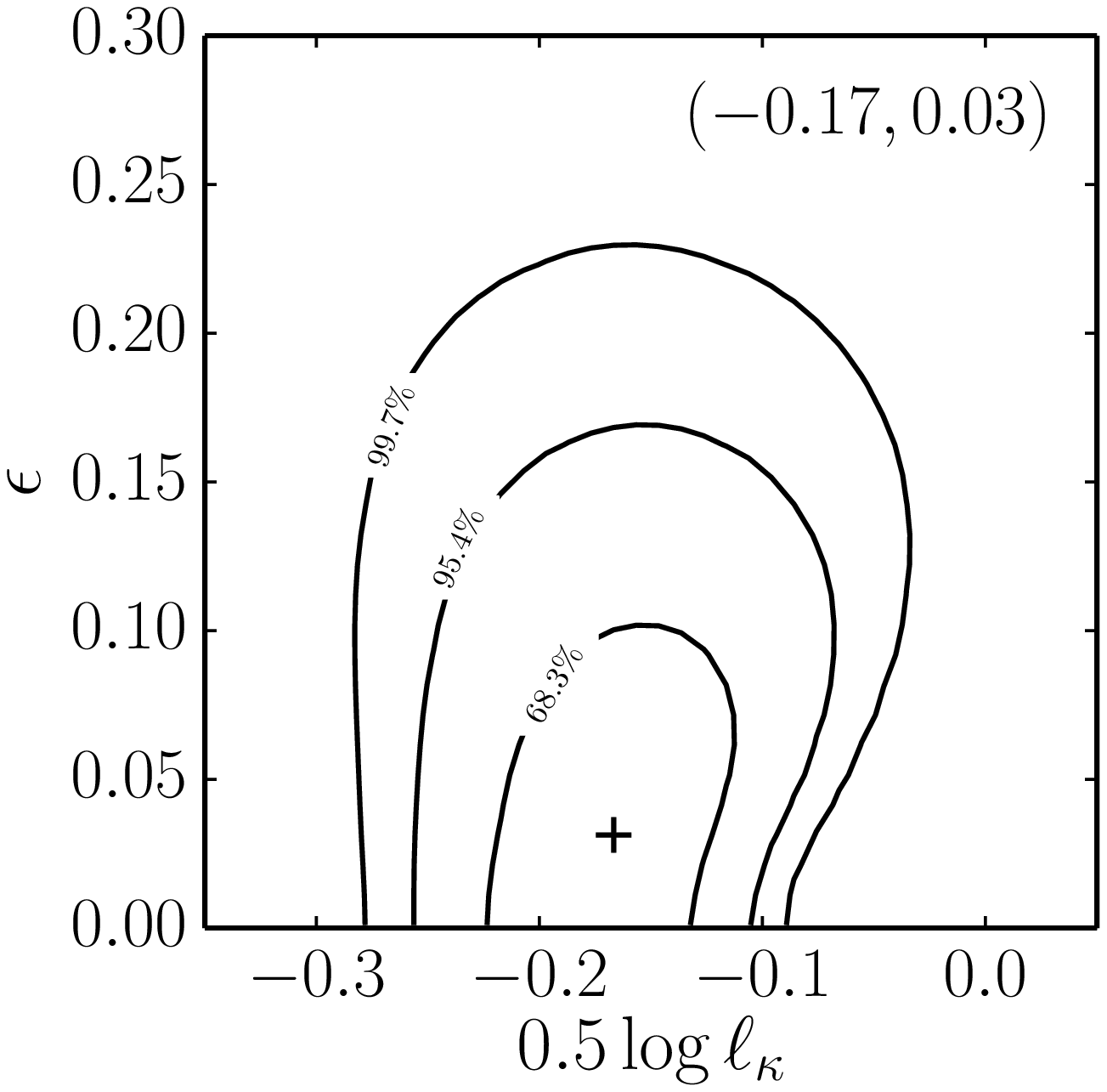}
\end{center}
\caption{\footnotesize
Significant level contours. {\it Left}, {\it middle} and {\it right} 
panels show ($\log\ellkappa-\beta$), ($\epsilon-\beta$) and ($\epsilon-\log\ellkappa$), 
respectively. Levels of significance are marked on the contour lines.
}
\label{kappa_factor}
\end{figure*}

\begin{figure*}[t!]
\begin{center}
\includegraphics[angle=0,width=0.9\textwidth]{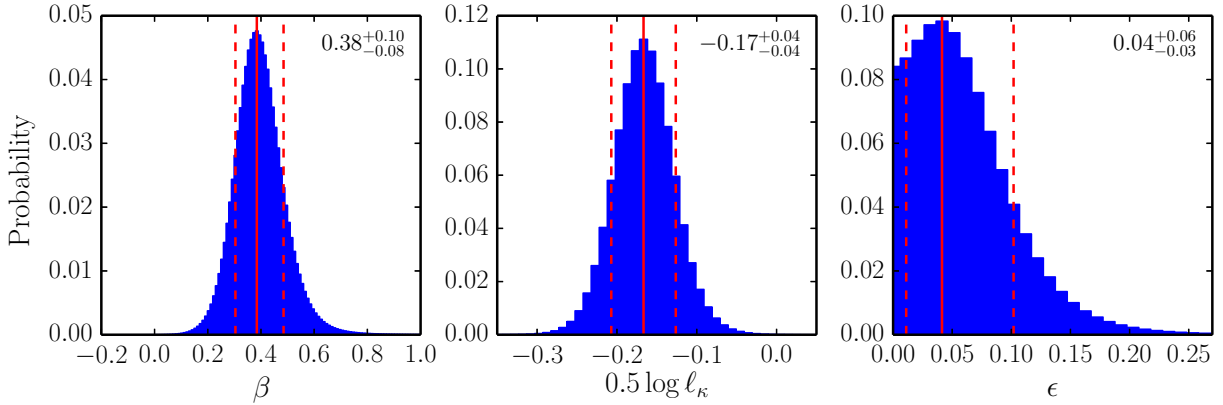}
\end{center}
\caption{\footnotesize
The calculated probabilities of $\log\ellkappa$(left), $\beta$ i(middle) and $\epsilon$ (right).
}
\label{hubble_diagram}
\end{figure*}

\section{The Cosmological distance of SEAMBHs}
\subsection{Basic equations}
To measure the distance from $\bhl$ using the new method we have to know the bolometric correction 
factor defined by 
$\kappabol=\bhl/L_{5100}$,
where $L_{5100}$ is the luminosity at 5100\AA. Ideally, this could be determined by direct observations 
and integration over the entire SEDs. As explained, this is not possible in the UV and EUV parts of the 
spectrum and hence this is normally replaced by an empirical estimate of $\kappabol$ such as the one 
calculated by Marconi et al. (2004). For geometrically thin disks, 
$\kappabol\propto (\mathdotM/\bhm)^{1/3}$ (Frank et al. 2002). As demonstrated by Netzer \& 
Trakhtenbrot (2014), the difference between these two approximations can be very large. The situation 
regarding slim disks  is even more problematic since models of such objects are rather uncertain which
reflects on the uncertainty in $\kappabol$.

The approach adopted here is to use the simple slim disk calculation (Abramowicz et al. 1988) 
to write a generic approximation for $\kappabol$ and use the observations to find the parameters 
in this equation. The expression is motivated by thin disk models and is given by 
\begin{equation}
 \kappabol=\kappa_0m_7^{-\beta} \, ,
\label{eq:kappabol}
\end{equation}
where $m_7=\bhm/10^7\sunm$. For the simplest slim disks $\kappa_0=40$ and $\beta=1/3$ (Mineshige 
et al 2000; see also Wang et al. 1999b, Shimura \& Manmoto 2003). 

Having defined the bolometric correction term, we define $F_{5100}$ as the measured 
$\lambda F_{\lambda}$  at $\lambda=5100(1+z)$\AA, after correcting for foreground extinction. 
Thus, 
$\Dbh=\left(\xi\bhl/4\pi \kappabol F_{5100}\right)^{1/2}$.
where $\xi$ is the radiation anisotropy factor which is a complex function of the 
disk geometry and inclination angle $i$ (e.g Madau 1988). We approximate this by 
$\xi = \cos i/0.5$ for a thin disk and by $\xi=1$ for a thick disk. For slim disks in type-I AGNs, 
$\cos i \simeq 0.75$ and $H/R\lesssim 1$ so  $\xi\approx 1$ is a good approximation.
We can now combine $\ell_0$, $\xi$ and $\kappa_0$ into one parameter,
\begin{equation}
\ell_{\kappa}=\xi\ell_0/\kappa_{40}, 
\label{eq:ell_kappa}
\end{equation}
where $\kappa_{40}=\kappa_0/40$. This result in the final expression for the distance:
\begin{equation}
\Dbh=250.3~\ell_{\kappa}^{1/2}m_7^{(1+\beta)/2}F_{11}^{-1/2}~{\rm Mpc} \, ,
\label{eq:dbh}
\end{equation}
where $F_{11}=F_{5100}/10^{-11}\ergs{\rm cm^{-2}}$. 

\begin{figure*}[t!]
\begin{center}
\includegraphics[angle=0,width=0.57\textwidth]{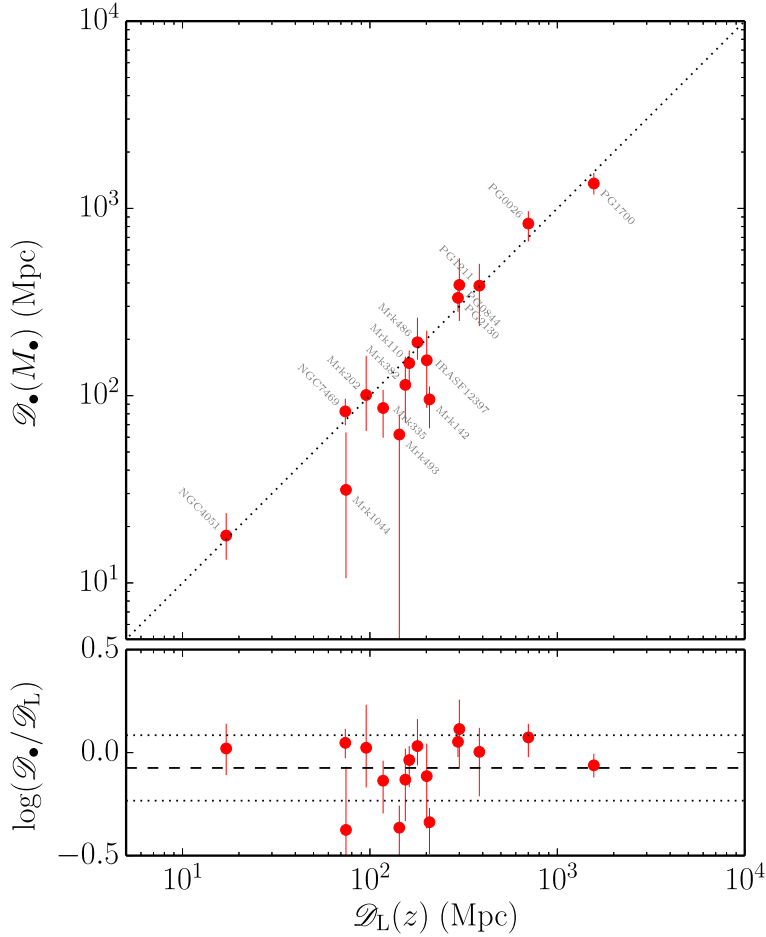} 
\end{center}
\caption{\footnotesize
Comparison of the newly derived distances, $\Dbh$, with those obtained from the standard
cosmological model. 
The dotted line in the upper panel is $\Dbh=\Dl$ and the dashed line in the bottom panel
indicates the mean values of $\log\Dbh^i/\Dl^i$ is $-0.06$, in our sample. 
The error bars are those listed in Table 4.
}
\label{hubble_diagram}
\end{figure*}
\vglue 0.5cm

\subsection{Error estimate and model uncertainties}
To understand the various uncertainties associated Eqn.~(\ref{eq:dbh}), we divide 
all the variables contributing to the error on the distance into three different groups.
The first group contains three measurable quantities, the continuum flux $F_{5100}$, 
the velocity $V_{\rm BLR}$ and time lag $\taublr$ that are used to calculate the 
mass (the uncertainty on the host galaxy flux is included in the uncertainty on $F_{5100}$). 
These uncertainties are listed in Table 4. The second group contains $\fblr$ which is not 
measured in individual sources but derived from the known properties of RM AGNs with host 
that show absorption lines. As explained, we use the Woo \etal\ (2013 and private 
communication) result of $\fblr = 1.00\pm 0.25$. In particular we use 
$\Delta\log \fblr=0.07$. 

The third group includes parameters that can in principle be obtained from the slim disk 
theory. For the simplified slim disk model used here these are 
$\ell_0$, $\xi$ and $\kappabol$ that are all absorbed into $\ell_\kappa$ 
(Eqn.~\ref{eq:ell_kappa}).
The anisotropy factor, $\xi$, is a complicated 
function of the disk geometry and include the effects of radiative transfer, inclination, 
reflection by the funnel walls, etc. As explained, the approximation adopted 
here is $\xi=1$ with an uncertainty similar to the uncertainty on $\cos i$.
For type-I AGNs, the inclination angle $i$ can change over
a small range of $\cos i=0.5-1$. With our chosen value of $\cos i=0.75$ we get
$\Delta \cos i/\cos i \simeq 0.33$ and $\Delta\log{\xi}= 0.07$.

The total uncertainty on a given $\Dbh$ is obtained by combining the uncertainties 
associated with the BH mass measurements and the flux, with those estimated for  
$\xi$ and $\fblr$ and are assumed to be the same for all objects. This should also 
be combined with the uncertainty on $\ellkappa$ which cannot be obtained
directly from the simple disk model and hence required different considerations. 
As we show below, the remaining uncertainties and the missing
constant of calibration, can be obtained by a comparison with the  observations.

\subsection{Distance Calibration}
We define $\dbh = \log \Dbh$ and write eqn.~\ref{eq:dbh} for  source $i$ as
\begin{equation}
\dbh = c_0 + \ell + (1+\beta)m_i - F_i + \epsilon_i,
\label{eqn:Dl}
\end{equation}
where $c_0=\log250.3=2.398$, $m_i = \frac{1}{2}\log m_7$, $F_i = \frac{1}{2}\log F_{11}$ 
and $\ell = \frac{1}{2}\log \lk$ (same for all sources). The new term, $\epsilon_i$, is 
the intrinsic scatter associated with the new method.
It represents the uncertainty related to additional physical parameters that were not 
included in our simple slim disk model (e.g., BH spin).
If such terms are important, they will introduce a large scatter which will dominate 
the uncertainty in $\dbh$. We assume that $\epsilon_i$ has a Gaussian distribution with 
a 1-$\sigma$ width of $\epsilon$.

The calibration of $\ell$ and $\beta$ is achieved by requiring that $\Dbh=\Dl$, where $\Dl$
is the luminosity distance derived from the standard cosmological
model. For low redshift sources, the uncertainty on $\Dl$ is equivalent to the uncertainty 
on $H_0$ which is less than 2\% (Freedman \& Madore 2013). This is significantly smaller 
than the uncertainties on $\Dbh$ (see Table 4)
and we do not include this in our Bayesian analysis.
The values of cosmological parameters chosen here are based on the recent Planck measurements
(Ade et al. 2013):  $H_0=67\,{\rm km~s^{-1}~Mpc^{-1}}$, $\Omega_{\rm M}=0.32$ and
$\Omega_{\Lambda}=0.68$ (note that the use of $H_0$ only is not enough since the redshift
range is $0-0.3$).

Using Bayes' theorem, the posterior probability is
\begin{equation}
p(\ell, \beta, \epsilon \mid \{m_i, F_i\}) \propto 
               p(\{m_i, F_i\} \mid \ell, \beta, \epsilon) \times p(\ell, \beta, \epsilon),
\end{equation}
where $p(\{m_i, F_i\} \mid \ell, \beta, \epsilon)$ is the likelihood function and 
$p(\ell, \beta, \epsilon)$ is the prior probability. Since the error distributions 
of $\{m_i,F_i\}$
are not Gaussian, we employ asymmetric Gaussians to approximate their distributions.
The function $p(\ell,\beta,\epsilon)$ is the prior probability, which is
assumed to be uniformly  distributed since we have no prior information about the
three parameters. In Appendix A, we derive the likelihood function.
The normalised posterior function is given by
\begin{equation}
p(\ell,\beta,\epsilon \mid \{m_i,F_i\})=\frac{p(\{m_i, F_i\} \mid \ell,\beta,\epsilon)}
{\iiint p(\{m_i, F_i\} \mid \ell,\beta,\epsilon)d\ell d\beta d\epsilon} \, .
\label{eqn:likelihood}
\end{equation}
The most likely values of $\log\ellkappa$ and $\beta$ with $\epsilon$ 
are located where the posterior function is maximized. The joint confidence region 
for $\log\ellkappa$ and $\beta$ is derived from the three-dimensional 
likelihood space via reducing the three-dimensional probability distribution function
to two dimensions
\begin{equation}
p(\ell,\beta|\{m_i,F_i\})=\int p(\ell,\beta,\epsilon|\{m_i,F_i\})d\epsilon  \, .
\end{equation}
This is plotted in Figure 4 left panel. The other two joint functions of confidence 
regions are given by $p(\beta,\epsilon|\{m_i,F_i\})$ and $p(\ell,\epsilon|\{m_i,F_i\})$, 
which can be obtained from Eq.~(\ref{eqn:likelihood}) by integrating $\ell$ and $\beta$, 
respectively. Employing the observational data given in Table 4, the joint functions of
confidence regions are plotted as contours in Figure 4. Integrating two of the three 
parameters in the joint confidence functions, we have the probability distributions 
of the three parameters in Figure 5. The most likelihood values and their uncertainties 
are given by
$$
\log \ell_{\kappa}=-0.34^{+0.08}_{-0.08};~~~ 
\beta=0.38^{+0.10}_{-0.08};~~~ 
\epsilon=0.04^{+0.06}_{-0.03}.
$$

The results obtained here show that the calibrated value of $\beta$ is consistent 
with $1/3$, which is the value derived for SS73-disks. This is related to the fact 
that in slim disks, most of the 5100\AA\ flux originates at a few $10^3r_g$, where 
photon trapping is not important. Therefore the bolometric correction factor follows 
a relation that is similar to the one obtained for thin accretion disks. The normalization 
factor $\ellkappa$ is also consistent with the simple model of slim disks. 
Most importantly, the value of $\epsilon$ which reflects several physical unknowns in 
the slim disk theory (spin, exact geometry, etc.)
is very small (0.04), well below the combined uncertainties of $\xi$ and $\fblr$. These 
physical unknowns cannot influence much the
accuracy of the new method confirming the suggestions that SEAMBHs can be used as cosmological 
distances indicators provided the measurement errors can be substantially reduced.

Using the values of $\ell_{\kappa}$ and $\beta$  found here, we can calculate the 
distances to all SEAMBHs listed in Table 4 and obtain the averaged residual distance 
($\log\Dbh^i/\Dl^i$) and its scatter in our sample. This number is $-0.06$ with a standard 
deviation of $\sigma_{\bullet}=0.14$.
Figure 6 shows the comparison of the new distance with $\Dl$ for all the sources in our
sample. The correlation in the upper panel is very close to a line of 1:1
and the plotted $\log\Dbh^i/\Dl^i$, in the (bottom panel, shows no dependence on the 
on the standard luminosity distance. In Figure 7 we examine the dependence of
$\log\Dbh^i/\Dl^i$ on BH mass. Here, again, there is no systematic deviation confirming the
usefulness of this technique over a large mass range. The error bars on the points in both
diagrams are the combination of all the uncertainties and are listed in 
the right column of Table 4.

\begin{figure}[t!]
\begin{center}
\includegraphics[angle=0,width=0.47\textwidth]{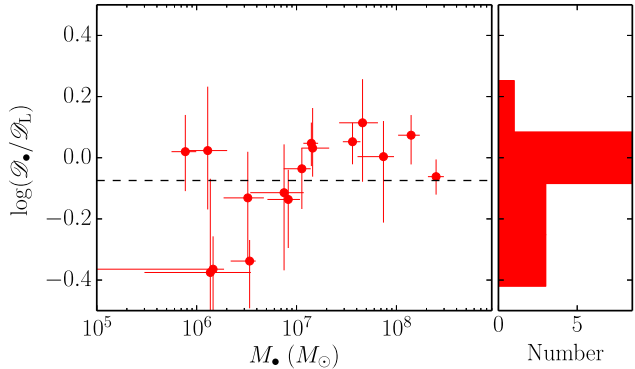} 
\end{center}
\caption{\footnotesize $\log\Dbh^i/\Dl^i$ vs. BH mass.
The dashed line is the mean value in our sample ($-0.06$).}
\label{hubble_diagram}
\end{figure}

\subsection{Implications for cosmology}
At $z=0.5-1$, the distance modular difference between an accelerating Universe and
a constant expanding universe is about 0.3\,mag (e.g., Riess et al.
1998). Using the scatter found here ($\sigma_{\bullet}$), we estimate that some 60-100 
SEAMBHs will be required, for the same redshift interval, to achieve this accuracy.
Obviously, a much larger number of sources will be required to constrain $w$ or its cosmological
evolution. It is too early to estimate this number given the limited number of sources (15) used here
and the potential improvements of the method once better slim disk models are available.

How likely is it to find a large number of SEAMBHs at high redshift and measure
their BH mass to the same level of accuracy achieved here?
Several recent papers describe searches for super-Eddington accreting massive black holes in 
large AGN samples like the SDSS (Nobuta \etal\, 2011; Trakhtenbrot \& Netzer 2012; Kelly \& 
Shen 2012; Netzer \& Trakhtenbrot 2014). Such studies use BH mass estimates based on the known 
$\rblr-L$ relation in combination with an empirical bolometric correction factor like the one 
derived by Marconi \etal\, (2004). All these studies show a large fraction of sources with 
$\mdot > 0.3$, i.e., likely to be powered by slim accretion disks. The recent work by Netzer 
\& Trakhtenbrot (2014) adopt an approach similar to the one used here. The paper assumes that 
all AGNs are powered by accretion disks and used the estimated $\bhm$ and $\Mdot$ to derive 
the value of $\mdot_{\rm min}$. This is then used to investigate the fraction of slim accretion 
disks in the AGN population. They found that at redshifts of about 0.7 and beyond, most of the 
SDSS detected AGNs contain slim accretion disks. Thus, there is no lack of suitable sources and 
the main obstacle is observational, to measure the mass of so many BHs by RM.

Several major improvements of the methodology  used here are around the corner promising to 
achieve higher accuracy in estimating  $\fblr$, $\xi$ and $\ell_{\kappa}$; the main contributers 
to the uncertainty. Estimates of $\fblr$ and $\xi$ can be improved by better modelling of the BLR
using methods like maximum entropy reconstruction space (Horne \etal\, 2004)
and/or the Markov chain Monte Carlo simulations (MCMC) (Pancoast \etal\, 2011, 2013; Li \etal\, 
2013). The uncertainties on $\ell_{\kappa}$ are due to theoretical limitations and incomplete
understanding of slim accretion disks. The main ingredient that requires improvement
is the treatment of the two-dimensional radiative transfer in the disk (Ohsuga \etal\, 2002, 
2005; Pountanen \etal\, 2007; Dotan \& Shaviv 2012; Yang \etal\, 2014). Better calculations 
will improve the estimates of $\eta$ in slim accretion disks and through it, the estimates of 
$\dotm$. As explained, the method does not depend much of such theoretical developments since 
$\ell_{\kappa}$ can be calibrated experimentally. 

\section{Summary and conclusions}
The first year of observations with the Lijiang 2.4-m telescope resulted in successful time 
lag and BH mass measurements of 8 radio-quiet narrow line Seyfert 1 galaxies. A conservative 
method 
to estimate the normalized accretion rate $\mdot$ show that 7 of those are powered by slim 
accretion disks and hence are SEAMBHs. The time lag in one of the 7 sources, Mrk\,493, is 
consistent with zero and hence removed from the rest of the analysis. Literature search for 
sources with RM-based BH mass measurements revealed 9 more sources that are SEAMBHs by our 
definition.

We presented  a novel method that can be used to infer cosmological distances using SEAMBHs. 
We tested the method on our sample of 15 SEAMBHs with $\dotm_{\rm min}>0.1$  and showed that 
it can be used to obtain cosmological distances. The key to the new method is the 
empirical calibration of the mean bolometric correction factor, and through it the saturated 
luminosity, by means of a Baysian analysis. The resulting bolometric correction is in agreement 
with the simple version of slim disk models and the intrinsic scatter of the distance
estimate is only 0.04 dex indicating that the unknown SEAMBH physics  is small and does
not prevent us from using the method.

For SEAMBHs of $10^7\sunm$, the intrinsic bolometric luminosities are few $\times 10^{45}\ergs$,
which is much higher than SNe-Ia (Leibundgut 2001). Furthermore, SN-based
cosmological distances at $z>1$ are rather uncertain due to small number statistics
and such objects are hard to find at high-$z$ (Hook 2012)
because of their faintness and/or the slow evolution of their progenitors (Kobayashi \etal\, 2009).
In contrast, the number density of massive, very luminous
fast accreting BHs is increasing with redshift (Trakhtenbrot \& Netzer 2012; Nobuta \etal\,
2012; Kelly \& Shen 2013; Netzer \& Trakhtenbrot 2014)
and the saturated luminosity of slim disks is nearly independent of redshift-related
factors such as the chemical composition of the accreted gas. Future RM experiments based on the 
\mgii\,$\lambda2798$\AA\, line, and perhaps other lines that will be proven 
to be good indicators for the motion of the 
BLR gas, can be used to measure BH mass in a large number of high-$z$ SEAMBHs. Such 
a campaign will require  large ground-based telescopes over a period of several years. The 
results can be used to measure BH mass and new cosmological distances and to explore the dynamics 
of the Universe in the era when gravity was the dominant, but not the only factor
affecting its expansion.

\acknowledgements The authors are very grateful to an anonymous referee for useful report that 
helped to improve the paper. We thank the staff of the Lijiang Station of the Yunnan
Observatory for their great help in making this project successful in the first
year of observations. We thank L. C. Ho and Y.-Y. Zhou for useful discussions. The research is 
supported by the Strategic Priority Research Program $-$ The Emergence of Cosmological Structures 
of the Chinese Academy of Sciences, Grant No. XDB09000000. This research is supported by the
NSFC through NSFC-11173023, -11133006, -11233003, and by the Israel-China ISF-NSFC grant 83/13.

\clearpage

\appendix
\section{Posterior function}
Considering the independent measurements of $\{m_i,F_i\}$,
we have the likelihood function
\begin{equation}
p(\{m_i, F_i\} \mid \ell, \beta, \epsilon) 
              = \prod_i p(m_i, F_i \mid \ell, \beta, \epsilon) \, 
\end{equation}
where the probability is given by 
\begin{equation}
\begin{array}{lll}
p(m_i, F_i \mid \ell,\beta,\epsilon) & = &\int p(m_i,F_i,\dbhu\mid \ell,\beta,\epsilon)\ud\dbhu \\
&= &\int p(m_i, F_i \mid \dbhu, \ell, \beta, \epsilon) p(\dbhu \mid \ell, \beta, \epsilon) \ud\dbhu \, .
\end{array}
\end{equation}
Here $\dbhu=d_{\rm L}^i-\epsilon_i$ is the actual distance of source $i$ which is not known.
thus we use the probability formulation of $p(X,Y)=p(X|Y)\times p(Y)$. The first term 
in the integral reads
\begin{equation}\begin{array}{lll}
p(m_i, F_i \mid \dbhu, \ell, \beta, \epsilon)
&=& \iint p(m_i, F_i, m, F \mid \dbhu, \ell, \beta, \epsilon) \ud m \ud F \\
&=& \iint p(m_i, F_i \mid m, F, \dbhu, \ell, \beta, \epsilon) \\
& & \times p(m \mid F, \dbhu, \ell, \beta, \epsilon) p(F \mid \dbhu, \ell, \beta, \epsilon)\ud m \ud F \\
&\propto& \iint p(m_i, F_i\mid m, F, \dbhu, \ell, \beta, \epsilon) \delta(m - m^{\prime}) \ud m \ud F,
\end{array}
\end{equation}
where $\delta(m-m^{\prime})$ is the $\delta$-function,
$p(F \mid \dbhu, \ell, \beta, \epsilon)$ is a uniform distribution, $m$ and $F$ 
are the actual black hole mass and flux, 
$m^{\prime} = (\dbhu - c_0 - \ell + F) / (1 + \beta)$. So
\begin{equation}
\begin{array}{lll}
p(m_i, F_i \mid \dbhu, \ell, \beta, \epsilon) 
&\propto& \iint p(m_i, F_i \mid m, F, \dbhu, \ell, \beta, \epsilon) \delta(m - m^{\prime}) \ud m \ud F, \\
&\propto& \int p(m_i, F_i\mid m^{\prime}, F, \dbhu, \ell, \beta, \epsilon) \ud F.
\end{array}
\end{equation}
The observational uncertainties on  $m_i$ and $F_i$ are independent hence, 
\begin{equation}
\begin{array}{lll} \label{eqn:pmf}
p(m_i, F_i \mid \dbhu, \ell, \beta, \epsilon)
&\propto& \int p(m_i, F_i \mid m^{\prime}, F, \dbhu, \ell, \beta, \epsilon) \ud F \\
&\propto& \int p(m_i\mid m^{\prime},F,\dbhu,\ell,\beta,\epsilon) 
               p(F_i\mid m^{\prime},F,\dbhu,\ell,\beta,\epsilon) \ud F.
\end{array}
\end{equation}
The likelihood function is
\begin{equation}
\begin{array}{lll} \label{eqn:pmf2}
p(\{m_i,F_i\}\mid \ell, \beta, \epsilon)
&\propto&\displaystyle
             \prod_i\iint p(m_i \mid m^{\prime},F,\dbhu,\ell,\beta,\epsilon) 
             p(F_i\mid m^{\prime},F,\dbhu,\ell,\beta,\epsilon) \\
&       &\times p(d_{\epsilon}\mid\ell,\beta,\epsilon) \ud F \ud \dbhu,
\end{array}
\end{equation}
where $p(m_i \mid m^{\prime}, F, \dbhu, \ell, \beta, \epsilon)$ and 
$ p(F_i \mid m^{\prime}, F, \dbhu, \ell, \beta, \epsilon)$ are given by the 
asymmetric Gaussian functions.

The probability of $p(\dbhu \mid \ell, \beta, \epsilon)$ in Eqn (A6) is
\begin{equation}
p(\dbhu \mid \ell, \beta, \epsilon) = \frac{1}{\sqrt{2\pi}\epsilon}
                                \exp\left[-\frac{(\dbhu-\dli)^2}{2\epsilon^2}\right],
\end{equation}
which is the normalized Gaussian with dispersion $\epsilon$.  

We approximate the errors of the measured properties by asymmetric Gaussian 
distributions,  
\begin{equation}
p(m_i\mid m^{\prime},F,\dbhu,\ell,\beta,\epsilon) 
           =\left\{\begin{array}{ll}\displaystyle
           \frac{1}{\sqrt{2\pi}\Delta_{m_i}}\exp\left[
           -\frac{(m_i-m^{\prime})^2}{2\Delta_{m_i}^2}
           \right]& ({\rm if}~ m_i >m^{\prime}),\\
           & \\
           \displaystyle
           \frac{1}{\sqrt{2\pi}\delta_{m_i}}\exp\left[
           -\frac{(m_i-m^{\prime})^2}{2\delta_{m_i}^2}
           \right] & ({\rm if}~m_i\le m^{\prime}),
           \end{array} \right.
\end{equation}
and
\begin{equation}
p(F_i\mid m^{\prime},F,\dbhu,\ell,\beta,\epsilon)  
           =\left\{\begin{array}{ll}\displaystyle
           \frac{1}{\sqrt{2\pi}\Delta_{F_i}}\exp\left[
           -\frac{(F_i-F)^2}{2\Delta_{F_i}^2}
           \right]& ({\rm if}~ F_i >F),\\
           & \\
           \displaystyle
           \frac{1}{\sqrt{2\pi}\delta_{F_i}}\exp\left[
           -\frac{(F_i-F)^2}{2\delta_{F_i}^2}
           \right] & ({\rm if}~F_i\le F),
           \end{array} \right.
\end{equation}
where $\Delta$ and $\delta$ are the upper and lower error bars of $m_i$ and $F_i$, respectively.

\clearpage

\end{document}